\documentclass[12pt]{article}
\usepackage{amssymb,amsmath}
\usepackage{latexsym}
\usepackage{graphicx}
\usepackage{psfrag}

\usepackage{booktabs}
\usepackage[margin=20pt,small]{caption}

\usepackage{bbm}
\usepackage{ifpdf}
\usepackage{longtable}
\usepackage[toc]{appendix}

\usepackage{color}
\usepackage{datetime}
\usepackage[
      colorlinks=false,
      linkcolor=darkblue,  
      urlcolor=blue,    
      filecolor=blue,     
      citecolor=red,
linktocpage=true,
      pdfstartview=FitV,
      bookmarksopen=true    
      ]{hyperref}

\ifpdf
\fi

\newcommand{\nosymmetry}{$\langle\mbox{No gauge symmetry}\rangle$}

\renewcommand{\theequation}{\arabic{section}.\arabic{equation}}

\def\coeff#1#2{\relax{\textstyle {#1 \over #2}}\displaystyle}

\def\IR{\mathbb{R}}

\def\cA{{\cal A}}
\def\cB{{\cal B}}

\def\cL{{\cal L}}

\def\cN{{\cal N}}
\def\cO{{\cal O}}
\def\cP{{\cal P}}

\def\cR{{\cal R}}
\def\cS{{\cal S}}

\def\cV{{\cal V}}
\def\cW{{\cal W}}

\def\Neql#1{{\cal N}\!=\!{#1}}
\def\eql{=}

\def\RR{\mathbb{R}}

\def\Tr{{\rm Tr}\,}

\def\labz{{z}}
\def\labw{{w}}

\definecolor{cardinal}{rgb}{0.6,0,0}
\definecolor{darkgreen}{rgb}{0,0.5,0}
\definecolor{golden}{rgb}{0.92, 0.7, 0}
\definecolor{midnight}{rgb}{0, 0, 0.5}
\definecolor{darkblue}{rgb}{0.2, 0, 0.8}

 \newcommand{\nc}{\newcommand}
\nc{\bea}{\begin{eqnarray}}
\nc{\eea}{\end{eqnarray}}
\nc{\be}{\begin{equation}}
\nc{\ee}{\end{equation}}
 \nc{\tphi}{\tilde{\phi}}
\nc{\non}{\nonumber}

\def\dx#1#2#3#4{{dx^#1\wedge dx^#2\wedge dx^#3\wedge dx^#4}}
\def\tau#1#2#3{{\,i \sigma^#1\otimes\sigma^#2\otimes\sigma^#3}}

\def\sotsotgp{SO(4)'}

\begin{document}  

\begin{titlepage}
 
\bigskip
\bigskip
\bigskip
\bigskip
\begin{center} 
{\Large \bf    New Supersymmetric and Stable, Non-Supersymmetric Phases  in Supergravity and Holographic Field Theory}

\bigskip
\bigskip 

{\bf Thomas Fischbacher${}^{(1)}$, 
 Krzysztof Pilch${}^{(2)}$,  and Nicholas P. Warner${}^{(2)}$ \\ }
\bigskip
${}^{(1)}$
University of Southampton\\
 School of Engineering Sciences\\
University Road, SO17 1BJ \\
Southampton, United Kingdom\\
\vskip 5mm
${}^{(2)}$
Department of Physics and Astronomy \\
University of Southern California \\
Los Angeles, CA 90089, USA  \\
\bigskip
t.fischbacher@soton.ac.uk,~pilch@usc.edu,~warner@usc.edu  \\
\end{center}

\begin{abstract}

  \noindent We establish analytically that the potential of $\Neql 8$
  supergravity in four dimensions has a new $\Neql 1$ supersymmetric
  critical point with $U(1) \times U(1)$ symmetry. We work within a 
  consistent $\Neql 1$ supersymmetric truncation and obtain the
  holographic flow to this new point from the $\Neql 8$ critical
  point.  The operators that drive the flow in the dual field theory
  are identified and it is suggested the new critical point might represent a new
  conformal phase in the holographic fermion droplet models with
  sixteen supersymmetries.  The flow also has $  \frac{c_{\rm IR} }{
    c_{\rm UV}} = {1 \over 2}$.  We  examine the stability of
  all 
  twenty known critical points and show that the $SO(3) \times
  SO(3)$ point is a perturbatively stable non-supersymmetric fixed
  point.
  We also locate and describe a novel critical point that also has
  $SO(3)\times SO(3)$ symmetry and is related to the previously known
  one by triality in a similar manner to the way that the $SO(7)^\pm$ 
  points are related to one  another.

\end{abstract}

\end{titlepage}


\tableofcontents

\section{Introduction}

In the early 1980's it was hoped that gauged $\Neql 8$ supergravity
would provide a theory of unified gauge interactions that was embedded
within a sensible theory of quantum gravity.  This hope went
unfulfilled for a number of fundamental reasons relating to the
problems of getting a viable vacuum with the appropriate chiral
fermions.  One of the issues was that the natural vacuum states all
seemed to have Planck-scale, negative cosmological constants.  Thus
$\Neql 8$ supergravity seemed to be an interesting, extremely symmetric
but rather rigid and unphysical theory.

The advent of holography changed this state of affairs precisely
because gauged $\Neql 8$ supergravity is the field theory of the
massless modes in the gravitational background generated by a stack of
$M2$ branes.  The $AdS$ vacua and related solutions could then be
reinterpreted in terms of conformal fixed points, perturbed fixed
points and flows in the holographically dual field theory on the $M2$
branes.  For the maximally symmetric ($\Neql 8$) vacuum, there is a
precise holographic dictionary that states that the fields of the
$\Neql 8$ supergravity are dual to relevant operators in the
energy-momentum tensor supermultiplet of the dual gauge theory.  In
particular, $\Neql 8$ supergravity contains fields that are dual to
fermion and boson bilinears in the $M$-brane field theory and so the
supergravity theory can be used to study holographic flows that are
driven by mass terms and vevs of such bilinears.  More recently, the
deeper understanding of the field theory on the $M2$ brane
\cite{Gustavsson:2007vu, Bagger:2007jr,Aharony:2008ug} enabled the
study of the gravity-gauge duality from both sides of the duality.
The interest in gauged $\Neql 8$ supergravity has also grown
considerably due to the extensive activity in $AdS/CMT$ where the
corresponding holographic field theories might be used to study
interesting, strongly coupled condensed matter systems in $(2+1)$
dimensions.

There have been two rather different approaches to the study of
$AdS/CMT$: The ``bottom-up'' (or ``phenomenological'') approach and the
``top-down'' approach.  In the former, the gravity dual of an
interesting condensed matter system is postulated {\it ab initio}
(see, for example, \cite{Gubser:2008px,Hartnoll:2008vx,
  Hartnoll:2008kx}), without using the more well-established
holographic field theories and their gravity duals.  In the latter
approach one tries to realize interesting phenomenological models
within theories that have well-established holographic dictionaries,
as one does in M-theory or IIB supergravity (see, for example,
\cite{Gubser:2008wz}-%
-\cite{Gauntlett:2009bh}). 
  It is very important to do this,
not only because of the dictionary, but also to make sure that the
complete holographic theory does not have other low-mass modes that
compromise or destroy the effect that one finds in the reduced or
effective field theory.  Indeed, there is at least one example of a
consistent, even supersymmetric theory with a (non-supersymmetric)
ground state that is completely stable within that consistent
truncation but is pathologically unstable within the complete
holographic theory \cite{Bobev:2010ib}.

Stability of $AdS$ vacua in supergravity has been well understood for
many years.  The supergravity potentials are generally unbounded below
and do not even have local minima but merely have critical points at
negative values of the potential.  The negative values of the
potential lead to anti-de Sitter ($AdS$) vacua that are expected to be
dual to non-trivial conformal fixed points in the field theory.  In
such a vacuum one can tolerate a certain amount of negative mass
because the gravitational back-reaction can stabilize the fluctuations
\cite{Breitenlohner:1982jf}.  To be
more precise, suppose that there is a scalar field, $\phi$, in
$d$-dimensions with a potential, $\cP(\phi)$, that has a critical
point at $\phi_0$. Taking the Lagrangian to be
\begin{equation}
e^{-1}{\cal L}\ ~=~ \coeff{1}{2}R - \coeff{1}{2} (\partial_\mu \phi)^2  ~-~ \cP(\phi )\,,
\label{scalLag}
\end{equation}
then the $AdS_d$ vacuum at $\phi_0$ is stable to quadratic
fluctuations if \cite{Breitenlohner:1982jf, Mezincescu:1984ev}
\begin{equation}
{(d-1)(d-2)\over 2}\,\left({\cP''(\phi_0)\over \cP(\phi_0)}\right)~\leq~ {(d-1)^2\over 4}\,.
\label{Bfcrit}
\end{equation}
For more general Lagrangians with more scalars and more complicated
kinetic terms, one expands quadratically, normalizes to the form of
(\ref{scalLag}) and then applies the Breitenlohner-Freedman (BF)
bound, (\ref{Bfcrit}), to the eigenvalues of the quadratic fluctuation
matrix for the potential.  In the holographically dual field theory,
violating the bound, (\ref{Bfcrit}), shows up as a manifest pathology:
The holographic dictionary shows that such perturbations are dual to
operators with complex conformal dimensions.
 
Supersymmetry guarantees stability through the usual energy bound
arguments applied to anti-de Sitter space \cite{Abbott:1981ff,
  Gibbons:1983aq}.  This implies complete classical and semi-classical
stability, and not merely to solutions based upon critical points.
Supersymmetric flow solutions are therefore completely stable but
imposing supersymmetry also significantly restricts the
physics. Indeed, in its simplest form, superfluidity and
superconductivity require the formation of a fermion condensate and 
at first sight it seems unclear whether such a condensation alone
could be rendered supersymmetric. It is also not immediately obvious
how a Fermi-sea would ever develop in a supersymmetric ground state.
On the other hand, we know from \cite{Lin:2004nb} that there are
holographic flows that have {\it sixteen} supersymmetries and whose
infra-red fixed point is described by free fermions and excitations of
the Fermi sea.  We will show that our new supersymmetric solutions and
flows are closely related to these holographic fermion theories and
that the new critical point might have the interpretation of a state
in this theory.

Rather remarkably, the non-supersymmetric $SO(3) \times
SO(3)$-invariant critical point discussed in \cite{Warner:1983du} has
not been subjected to a full stability analysis.  This critical point
is also part of the $\Neql 5$ truncation of $\Neql 8$ supergravity,
where it has an $SO(3)$ invariance.  In this setting it was  discussed
in \cite{Warner:Thesis, Gibbons:1983aq} and was shown to be
perturbatively stable within the $\Neql 5$ theory
\cite{Breitenlohner:1982jf,Warner:1983vz}.  Moreover a positive mass
theorem was established within the $\Neql 5$ theory
\cite{Boucher:1984yx} for this critical point and so it was shown to
be completely semi-classically stable.  This is not sufficient to
guarantee perturbative stability within the full $\Neql 8$ theory
because other fields could destabilize the vacuum, as in
\cite{Bobev:2010ib}, but we will show here that all the scalar
fluctuations around this point satisfy the BF bound and so this point
is the first known, perturbatively stable, non-supersymmetric $AdS$
vacuum for the $\Neql 8$ theory.\footnote{This finally and definitively
  invalidates the ``folk theorem'' that suggest that only supersymmetric critical
  points are stable in maximal supergravity  theories in more than three dimensions. 
  BF stability without supersymmetry  in a maximally 
  supersymmetric theory was first observed in the three-dimensional,  $SO(8)\times SO(8)$ 
  gauged $\Neql 16$ Chern-Simons model:  See
 formula~$(5.8)$ in~\cite{Fischbacher:2002fx}.}  
It would be interesting to see if the arguments of
\cite{Boucher:1984yx} could be extended to the complete $\Neql 8$
theory to establish complete semi-classical stability.

As a result of all these factors it was, and continues to be, very
important to classify the solutions of gauged $\Neql 8$ supergravity
in four dimensions and, most particularly, find the critical points of
the scalar potential of that theory and look at their stability. A
systematic analytic method for computing the potential was
developed a long time ago \cite{Warner:1983du} and was exploited in
\cite {Warner:1983vz} to obtain a variety of interesting new
supersymmetric and non-supersymmetric vacua.  The limitation of this
approach is that analytic computations generally require a fairly high
level of symmetry in order for the explicit computations to be
possible.  Thus, for over 25 years, the only known critical points of
the $\Neql 8$ potential had at least $SU(3)$ or $SO(3) \times SO(3)$
symmetry.  On the other hand, a numerical approach to the problem that
utilizes sensitivity back-propagation to obtain fast high quality
gradients recently has revealed that this potential has many more 
critical points \cite{Fischbacher:2009cj,
  Fischbacher:2010ki} with relatively low levels of symmetry.  The
fourteen new points discovered in \cite{Fischbacher:2009cj} have at
most $U(1) \times U(1)$ symmetry and eight of them have no residual
continuous symmetry at all.  Given this low level of symmetry, it is
highly unlikely, and probably impossible for analytic searches to have
discovered them.  (Indeed, in  Appendix~\ref{numlocations} we give a first 
glimpse of the difficulty of this problem.) Moreover, even if one knows exactly
where these critical points lie, the scalar expectation values are
generically so complicated that the $E_{7(7)}$ matrix has a
characteristic polynomial of very high degree and this renders
analytic exponentiation computationally out of reach.

One of the important discoveries in \cite{Fischbacher:2009cj} was the
possibility of a new $\Neql 1$ supersymmetric critical point with
$U(1) \times U(1)$ symmetry and with $\cP = -12 g^2$.  This point is
potentially very interesting for several reasons.  First, it is a new
supersymmetric point.  Second, the $U(1) \times U(1)$ symmetry
commutes with an $SO(4) \subset SO(8)$ gauge symmetry and this means
that there are a rich collection of possible gauge fields and hence
chemical potentials that could induce a flow to this critical point.
Moreover, the value of the cosmological constant (dictated by the
value of $\cP$) is less than that of all the critical points that have, at least, 
an $SU(3)$ invariance \cite{Warner:1983vz}.   The $c$-theorem
\cite{Freedman:1999gp} implies that the cosmological constant must
decrease along holographic flows thus interesting, but unstable, flows
in $AdS/CMT$, like some of those considered in \cite{Gauntlett:2009dn,
  Gauntlett:2009bh}, might be arranged to ultimately flow to this new
$\Neql 1$ supersymmetric critical point.  Finally, as we will discuss,
this new supergravity phase may well have an interesting holographic
interpretation in terms of a new phase of the fermion droplet model
considered in \cite{Lin:2004nb}.

Fortunately, since one now knows where to look, this point can be
constructed analytically within a consistent truncation of the $\Neql
8$ theory and in this paper we construct this truncation, give an
analytic description of the point showing that it does indeed have
precisely the symmetries suggested in \cite{Fischbacher:2009cj}.  We
also give a superpotential on the truncated sector and obtain the
supergravity flow to the new $\Neql 1$ point from the $\Neql 8$ point.

In Section \ref{contrunc} we discuss successive consistent truncations
of the $\Neql 8$ theory down to a simple $\Neql 1$ theory that we will
then analyze in detail.  We include some intermediate $\Neql 4$
and $\Neql 2$ intermediate truncations that not only make the ultimate
truncation clearer but should also prove valuable if one
wishes to include vector multiplets so as to include holographic
chemical potentials.  In Section \ref{susyflows} we construct the
scalar potential and superpotential of this $\Neql 1$ theory and show
that the potential has critical points corresponding to the
non-supersymmetric $SO(3) \times SO(3)$-invariant point and to the new
$\Neql 1$ supersymmetric point found in \cite{Fischbacher:2009cj}.
The superpotential only has the maximally supersymmetric critical
point and the $\Neql 1$ supersymmetric critical point.  We also obtain
the holographic flow equations for this superpotential and exhibit the
$\Neql 1$ supersymmetric flow from the $\Neql 8$ point to the $\Neql
1$ point.

In Section \ref{holFT} we consider the UV asymptotics of the
holographic flow and identify the non-normalizable and normalizable
fields involved.  It turns out that the only non-normalizable field
involved is the one that was extensively studied in \cite{Pope:2003jp,
  Bena:2004jw,Lin:2004nb} and whose holographic field theory may be
described, at least in the infra-red,  in terms of free fermions in $(1+1)$-dimensions.  
The flow we consider here is rather more complicated and involves modes that
break a lot of symmetry and that were not considered in
\cite{Pope:2003jp, Bena:2004jw,Lin:2004nb}, however, these extra modes
are {\it normalizable} and this strongly suggests that the flow here
might represent some exotic supersymmetric state of the fermion
droplet model defined in \cite{Lin:2004nb}.

In Section \ref{masses2} we examine the $SO(3) \times SO(3)$ invariant
point in more detail, show that it is stable and discuss possible
flows to this point. In Section~\ref{newso3point}, we discuss the analytic
properties of a \emph{novel} and different $SO(3)\times SO(3)$ symmetric critical
point that is unstable. In Section \ref{OtherPts} we consider all the
other non-supersymmetric critical points and discuss their mass
spectrum.  Out of the eighteen known, non-supersymmetric critical
points, it is only the $SO(3) \times SO(3)$-invariant point with
$\cP=-14g^2$ that satisfies the BF bound and is thus perturbatively
stable.  We also discuss the prospects of finding more critical points
and describe why we believe that there are probably many more critical
points with very low levels of symmetry.  We also describe why such
points are almost certainly outside the reach of analytic algorithms
but are likely to be accessible using numerical searches.  Finally, in
Section \ref{Concs}, we make some remarks about the broader
implications of our work.

\section{Consistent truncations}
\label{contrunc}

To examine the new supersymmetric point, we will describe it as part
of an $\Neql 1$ supersymmetric truncation of the $\Neql 8$ theory.  We
will, however, arrive at this $\Neql 1$ truncation successively by
reducing to sectors that are invariant under progressively more
symmetry.  Indeed, we will first pass to an $\Neql 4$ theory and then
show that the $\Neql 1$ theory lies in the common overlap of two
different $\Neql 2$ theories.  We anticipate that these intermediate
$\Neql 2$ truncations will be useful when one wishes to study more
general supergravity flows that involve vector fields and induce
chemical potentials in the holographic field theory.

\subsection{An intermediate $\Neql 4$ theory}
\label{Neql4sector}

Since the critical point we seek is invariant under under the two
rotations, ${\cal R}_{36}$ and ${\cal R}_{45}$, that generate an
$SO(2)^2$ subgroup of $SO(8)$
\cite{Fischbacher:2009cj,Fischbacher:2010ki}, the obvious first step
is to consider the sector of the entire $\Neql 8$ theory that is
invariant under this symmetry.

These rotations preserve four supersymmetries, $\varepsilon^1,
\varepsilon^2,\varepsilon^7$ and $\varepsilon^8$, and so the result
must be $\Neql 4$ supersymmetric.  Indeed it is $\Neql 4$ supergravity
coupled to two $\Neql 4$ vector multiplets: the six vector fields in
the graviton multiplet are $A_\mu^{IJ}$, $I,J \in \{1,2,7,8\}$ and the
two vector multiplets are generated by $A_\mu^{36}$ and $A_\mu^{45}$.

To isolate the scalar manifold, consider the $SO(4)_A \times SO(4)_B$
subgroup of $SO(8)$ where $SO(4)_A$ acts on indices $(1278)$ and
$SO(4)_B$ acts on indices $(3456)$.  Inside $E_{7(7)}$,  $SO(4)_B$
commutes with $SU(4) \times SL(2,\IR)$ where the $SU(4) \subset SU(8)$ acts on the
$(1278)$ and the non-compact generators of the $SL(2,\IR)$ can be
represented by the self-dual form:
\begin{equation}
\Upsilon_0 ~\equiv~    (\labw_0 \, dx_1 \wedge dx_2 \wedge dx_7 \wedge dx_8  +  \overline \labw_0 \, dx_3 \wedge dx_4 \wedge dx_5 \wedge dx_6)\,,
\label{SL2a}
\end{equation}
where $\labw_0$ is a complex number.  This $SL(2,\IR)$ defines the
complex scalar of the $\Neql 4$ graviton supermultiplet.  The
commutant of the $SO(2)^2$ generated by ${\cal R}_{36}$ and ${\cal
  R}_{45}$ in $E_{7(7)}$ is simply $SO(6,2)\times SL(2,\IR)$ where
$SU(4) \cong SO(6)$ is embedded in the obvious manner.  The
non-compact generators of the $SO(6,2)$ can be represented by the
self-dual forms:
\begin{equation}\label{Liescalars}
\begin{split}
\Upsilon_1 & ~\equiv~    (\labz_1 \, dx_1 \wedge dx_3 \wedge dx_6 \wedge dx_7  -  \bar \labz_1 \, dx_2 \wedge dx_4 \wedge dx_5 \wedge dx_8)\,,    \\
\Upsilon_2 & ~\equiv~    (\labz_2 \, dx_1 \wedge dx_4 \wedge dx_5 \wedge dx_7  -  \bar \labz_2 \, dx_2 \wedge dx_3 \wedge dx_6 \wedge dx_8) \,,   \\
\Upsilon_3 & ~\equiv~    (\labz_3 \, dx_1 \wedge dx_4 \wedge dx_5 \wedge dx_8  +  \bar \labz_3 \, dx_2 \wedge dx_3 \wedge dx_6 \wedge dx_7) \,,   \\
\Upsilon_4 & ~\equiv~    (\labz_4 \, dx_1 \wedge dx_3 \wedge dx_6 \wedge dx_8  +  \bar \labz_4 \, dx_2 \wedge dx_4 \wedge dx_5 \wedge dx_7) \,,  \\
\Upsilon_5 & ~\equiv~    (\labz_5 \, dx_1 \wedge dx_2 \wedge dx_3 \wedge dx_6  +  \bar \labz_5 \, dx_4 \wedge dx_5 \wedge dx_7 \wedge dx_8) \,,   \\
\Upsilon_6 & ~\equiv~    (\labz_6 \, dx_1 \wedge dx_2 \wedge dx_4 \wedge dx_5  +  \bar \labz_6 \, dx_3 \wedge dx_6 \wedge dx_7 \wedge dx_8) \,. 
\end{split} 
\end{equation}
This defines the twelve scalars in the two $\Neql 4$ vector multiplets. 

\subsection{Two intermediate  $\Neql 2$ theories}
\label{Neql2sectors}

One can now use some discrete subgroups of $SO(8)$:
\begin{eqnarray}
g_1: &&  (x_1, x_2, x_3, x_4) ~ \leftrightarrow~  - (x_1, x_2, x_3, x_4)  \,,   \\
g_2: &&  (x_1, x_3, x_6, x_7) ~ \leftrightarrow~  (-x_1, x_4, x_5, x_8) \,,
\end{eqnarray}
with the remaining $x_j$ invariant.  Note that these transformations
have determinant one and so live in $SO(8)$.  More importantly, these
discrete symmetries leave invariant the scalars that define the new
$\Neql 1$ critical point.

The two intermediate $\Neql 2$ theories are defined by requiring
invariance under either $g_1$ or $g_2$ separately.

Invariance under $g_1$ leaves the two supersymmetries, $\varepsilon^7$
and $\varepsilon^8$ and only two vector fields: $A_\mu^{78}$ and
$A_\mu^{12}$. From the supersymmetries it is evident that $A_\mu^{78}$
must lie in the $\Neql 2$  graviton multiplet while $A_\mu^{12}$ must 
generate an $\Neql 2$ vector multiplet. The scalars in the $SL(2,\IR)$ 
defined by (\ref{SL2a}) belong to this vector multiplet.  The remaining fields
then make up two $\Neql 2$ hypermultiplets and  these contain the
four complex $g_1$-invariant scalars in (\ref{Liescalars})
parametrized by $z_1, z_2, z_3$ and $z_4$ in (\ref{Liescalars}).  Thus the 
complete scalar manifold can be described in terms of the coset:
\begin{equation}
{SL(2,\IR) \over  SO(2)}  \times  {SU(2,2) \over  SU(2) \times SU(2) \times U(1)}  \,.
\label{N2coseta}
\end{equation}

Invariance under $g_2$ leaves the two supersymmetries, $\varepsilon^2$
and $\varepsilon^7 + \varepsilon^8$ and three vector fields:
$B_\mu^{(0)} \equiv A_\mu^{27} + A_\mu^{28}$, $B_\mu^{(1)} \equiv
A_\mu^{17} - A_\mu^{18}$ and $B_\mu^{(2)} \equiv A_\mu^{36} +
A_\mu^{45}$. From the supersymmetries it is evident that $B_\mu^{(0)}$
must lie in the graviton multiplet while $B_\mu^{(1)}$ and
$B_\mu^{(2)}$ generate two $\Neql 2$ vector multiplets. Again, the
scalars in the $SL(2,\IR)$ defined by (\ref{SL2a}) belong to one of
these vector multiplets.  There are three complex $g_2$-invariant
scalars that arise from (\ref{Liescalars}) by taking $z_3 = -z_1$,
$z_4 = -z_2$, $z_6 = -z_5$.  One of these belongs to a vector
multiplet while the remainder constitute an $\Neql 2$ hypermultiplet.
Thus the complete scalar manifold can  be described in terms of
the coset:
\begin{equation}
{SL(2,\IR) \over  SO(2)}  \times {SL(2,\IR) \over  SO(2)}  \times  {SU(2,1) \over  SU(2)   \times U(1)}  \,.
\label{N2cosetb}
\end{equation}
%

\subsection{Defining and parametrizing the $\Neql 1$ theory}
\label{Neql1sector}

Requiring invariance under both $g_1$ and $g_2$ leaves only one
supersymmetry, $\varepsilon^7 + \varepsilon^8$, and projects out all
the vector fields.  The $SL(2,\IR)$ of the $\Neql 4$ graviton
multiplet survives while, in (\ref{Liescalars}), one must set $\labz_5
= \labz_6 =0$ and $\labz_3 = -\labz_1$ and $\labz_4 = -\labz_2$.  This
reduces $SO(6,2)$ to $SO(2,2) \cong SL(2,\IR)\times SL(2,\IR)$ and so
the scalar coset is now:
\begin{equation}
{SL(2,\IR) \over  SO(2)}  \times {SL(2,\IR) \over  SO(2)}  \times   {SL(2,\IR) \over  SO(2)} \,.
\label{N2cosetc}
\end{equation}
These simply form three $\Neql 1$ scalar multiplets with the
$g_1$-invariant and $g_2$-invariant fermions:
\begin{equation}
\chi^{127}  -  \chi^{128} \,, \qquad \chi^{136}  - \chi^{145} \,, \qquad   \chi^{236}  + \chi^{245}   \,.
\end{equation}
We are thus dealing with $\Neql 1$ supergravity coupled to three
scalar multiplets.

We will work with this theory and we will use the following explicit
parametrization of the $SL(2,\IR)^3$:
\begin{align}
\Sigma_0 &~\equiv~    (\labw_0 \, dx_1 \wedge dx_2 \wedge dx_7 \wedge dx_8  +  \bar \labw_0 \,  dx_3 \wedge dx_4 \wedge dx_5 \wedge dx_6 )\,, \\[6 pt]
\Sigma_1 &~\equiv~   (\labw_1+ \labw_2) \,  ( dx_1 \wedge dx_3 \wedge dx_6 \wedge dx_7 -  dx_1 \wedge dx_4 \wedge dx_5 \wedge dx_8 )  \nonumber \\  
& \qquad   -   (\overline \labw_1+ \overline \labw_2) \, ( dx_2 \wedge dx_4 \wedge dx_5 \wedge dx_8  + dx_2 \wedge dx_3 \wedge dx_6 \wedge dx_7) \,,  \\[6 pt]
\Sigma_2  &~\equiv~     (\labw_1- \labw_2) \, (dx_1 \wedge dx_4 \wedge dx_5 \wedge dx_7  -  dx_1 \wedge dx_3 \wedge dx_6 \wedge dx_8) \nonumber \\ 
& \qquad  -   (\overline \labw_1 - \overline \labw_2) \, (dx_2 \wedge dx_3 \wedge dx_6 \wedge dx_8 +   dx_2 \wedge dx_4 \wedge dx_5 \wedge dx_7)   \,,
\end{align}
where we have set $\labz_1 = (\labw_1+ \labw_2)$ and $\labz_2 =
(\labw_1- \labw_2)$.  The three commuting $SL(2,\IR)$'s are then
parametrized by each of the $\labw$'s.  We will use the polar
parametrization with:
\begin{equation}
\labw_0 ~=~ \frac{1}{2 \sqrt{2}} \, \lambda_0 \, e^{ i \varphi_0} \,, \qquad \labw_1 ~=~\frac{1}{4 \sqrt{2}} \,  \lambda_1 \, e^{i \varphi_1} \,, \qquad \labw_2 ~=~ \frac{1}{4 \sqrt{2}} \, \lambda_2 \, e^{i \varphi_2 } \,, 
\end{equation}
and then use the exponentiated coordinates:
\begin{equation}
\zeta_0 ~=~ \tanh \big(  \coeff{1}{\sqrt{2}} \, \lambda_0 \big  ) \, e^{- i \varphi_0} \,, \qquad \zeta_1 ~=~ \tanh \big(\coeff{1}{2}\, \lambda_1 \big  ) \, \, e^{- i \varphi_1} \,, \qquad  \zeta_2 ~=~ \tanh \big(\coeff{1}{2}\, \lambda_2 \big  ) \, e^{+ i \varphi_2 } \,.
\end{equation}
%

\subsection{The scalar action}
\label{scalact}

The kinetic terms of the $\Neql 8$ theory reduce to precisely what one
would expect of an $(SL(2,\IR))^3$ coset:
\begin{equation}
\label{sckin}
\begin{split}
e^{-1}\cL_{kin.}  &~=~   -{1\over 96} \cA_\mu{}^{ijkl}\cA^\mu{}_{ijkl}  \\[6 pt]
 &~=~  \frac{1}{2} \,  \sum_{j =0}^2 \, \big(\partial_\mu \lambda_j \big)^2  ~+~ \frac{1}{4} \,   \sinh^2(\sqrt{2} \lambda_0) (\partial_\mu \varphi_0)^2~+~ \frac{1}{2} \,   \sum_{j =1}^2 \sinh^2( \lambda_j) (\partial_\mu \varphi_j)^2 \\[6 pt]
 &~=~   \frac{\big|\partial_\mu \zeta_0 \big|^2} {(1-|\zeta_0|^2)^2} ~+~ 2  \sum_{j =1}^2 \, \frac{\big|\partial_\mu \zeta_j \big|^2} {(1-|\zeta_j|^2)^2}  \,,
\end{split}
\end{equation}
where the normalizations are determined by the embedding indices of
the $SL(2, \IR)$'s inside $E_{7(7)}$.

The scalar potential in the $\Neql 8$ theory is given by:
\begin{equation}
\label{scalpot}
\begin{split}
\cP &~=~ -g^2\big(\coeff{3}{4}\left|A_1{}^{ij}\right|^2-\coeff{1}{24}\left|A_{2i}{}^{jkl}\right|^2\big) \\[6 pt]
&~=~  - g^2\Big(\,  4\, \cosh  \lambda_1 \, \cosh \lambda_2    + 2\, \cosh ( \sqrt{2}  \lambda_0) \\[6 pt]
& \qquad \quad - \coeff{1}{2}\, \sinh^2  \lambda_1 \,  \sinh^2  \lambda_2 \, \big[\cosh ( \sqrt{2}  \lambda_0)  \,(1 - \cos (2\varphi_1) \,\cos (2\varphi_2))\\[6 pt]
&  \qquad\qquad \qquad   \qquad \qquad -  \sinh ( \sqrt{2}  \lambda_0) (  \cos (2\varphi_1)- \cos (2 \varphi_2))  \,  \cos \varphi_0 \big] \Big)\,,
\end{split}
\end{equation}
This potential has exactly three inequivalent critical points.  The maximally
supersymmetric vacuum:
\begin{equation}
 \lambda_0 ~=~   \lambda_1 ~=~  \lambda_2~=~ 0\,; \qquad   \cP ~=~ - 6\, g^2  \,.
\end{equation}

The other two points are:
\begin{equation}
 \lambda_0 = 0\,, \ \  \lambda_1 =  \lambda_2 =  \pm \log(2+ \sqrt{5}) \,;  \ \   \varphi_1 =  \varphi_2 =  \frac{\pi}{4}  \,,  \qquad   \cP ~=~ - 14\, g^2  \,,
 \label{SO3pt}
\end{equation}
and
\begin{equation}\label{newsusypt}
\begin{split}
  \coeff{1}{\sqrt{2}} \lambda_0 ~=~ \log \Big[ \frac{1}{\sqrt{2}} &\,(\sqrt{3} \pm 1) \Big] \approx  \pm 0.65848  \,, \quad  \lambda_1 =  \lambda_2 =  \log \big[  \sqrt{3}  \pm  \sqrt{2}  \big]  \approx \pm 1.146216\,, \\[6 pt]
 \varphi_0  &~=~  \pm \frac{\pi}{2}  \,, \ \    \varphi_1 ~=~  \varphi_2 ~=~  \frac{\pi}{4}  \,;  \qquad  \qquad   \cP ~=~ - 12\, g^2  \,.
\end{split}
 \end{equation}
In terms of the complex coordinates, these are:
\begin{equation}
\zeta_0  = 0\,, \qquad \zeta_1 =    \pm\, \frac{1}{2} \, ( \sqrt{5}-1)  \,(1 -  i)\,,  \qquad  \zeta_2 =    \pm\, \frac{1}{2} \, ( \sqrt{5}-1)  \,(1 + i)\,,   \qquad   \cP ~=~ - 14\, g^2  \,,
 \label{SO3ptcplx}
\end{equation}
and
\begin{equation}
\zeta_0 =  \mp \frac{1}{ \sqrt{3}} \, i \,, \quad \zeta_1 =    \pm \frac{1}{2}  \,(\sqrt{3}-1) \,(1 -  i)\,,  \quad  \zeta_2 =    \pm \frac{1}{2}  \,(\sqrt{3}-1) \,(1 + i)\,,   \quad   \cP ~=~ - 12\, g^2  \,.
 \label{newsusyptcplx}
\end{equation}

\begin{figure}[t]
\centering
    \includegraphics[width=7.5cm]{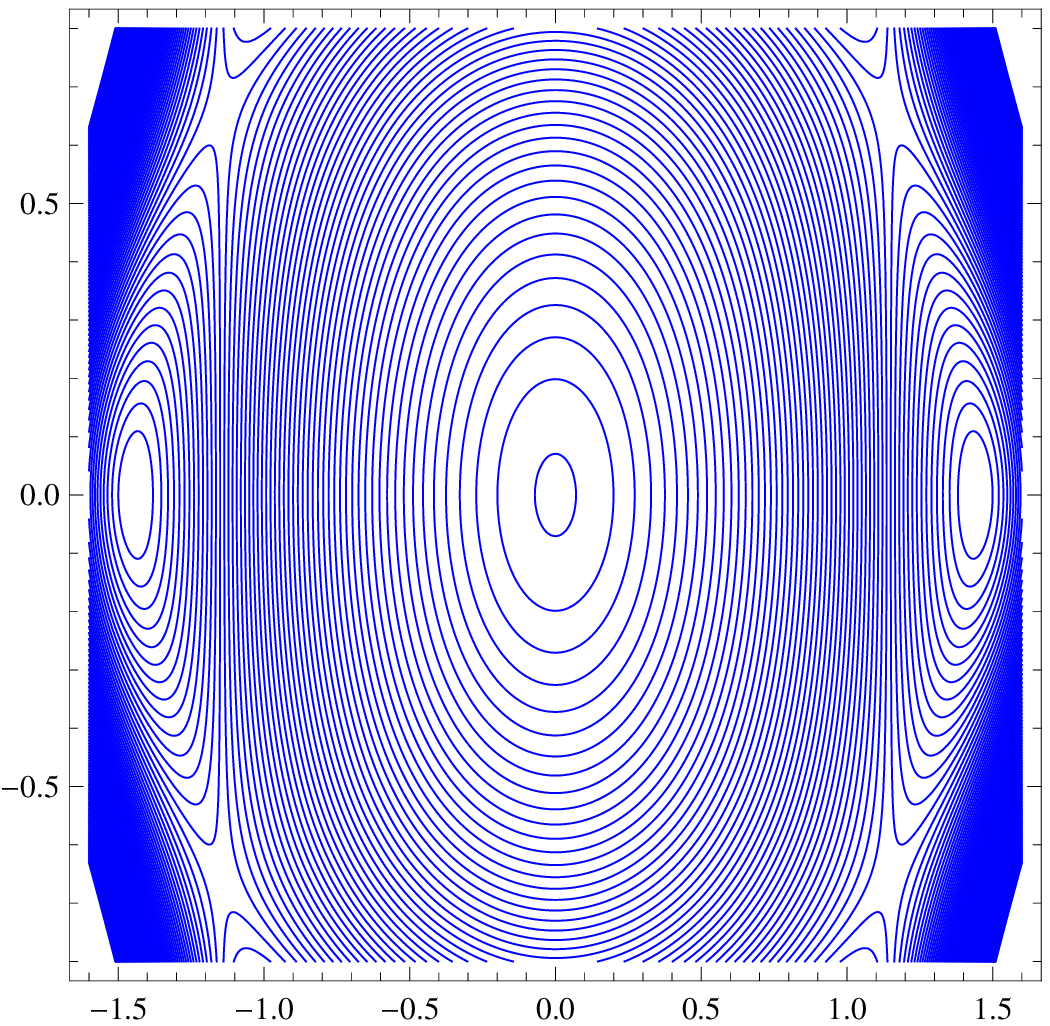}
    \hskip 1.0cm
    \includegraphics[width=7.5cm]{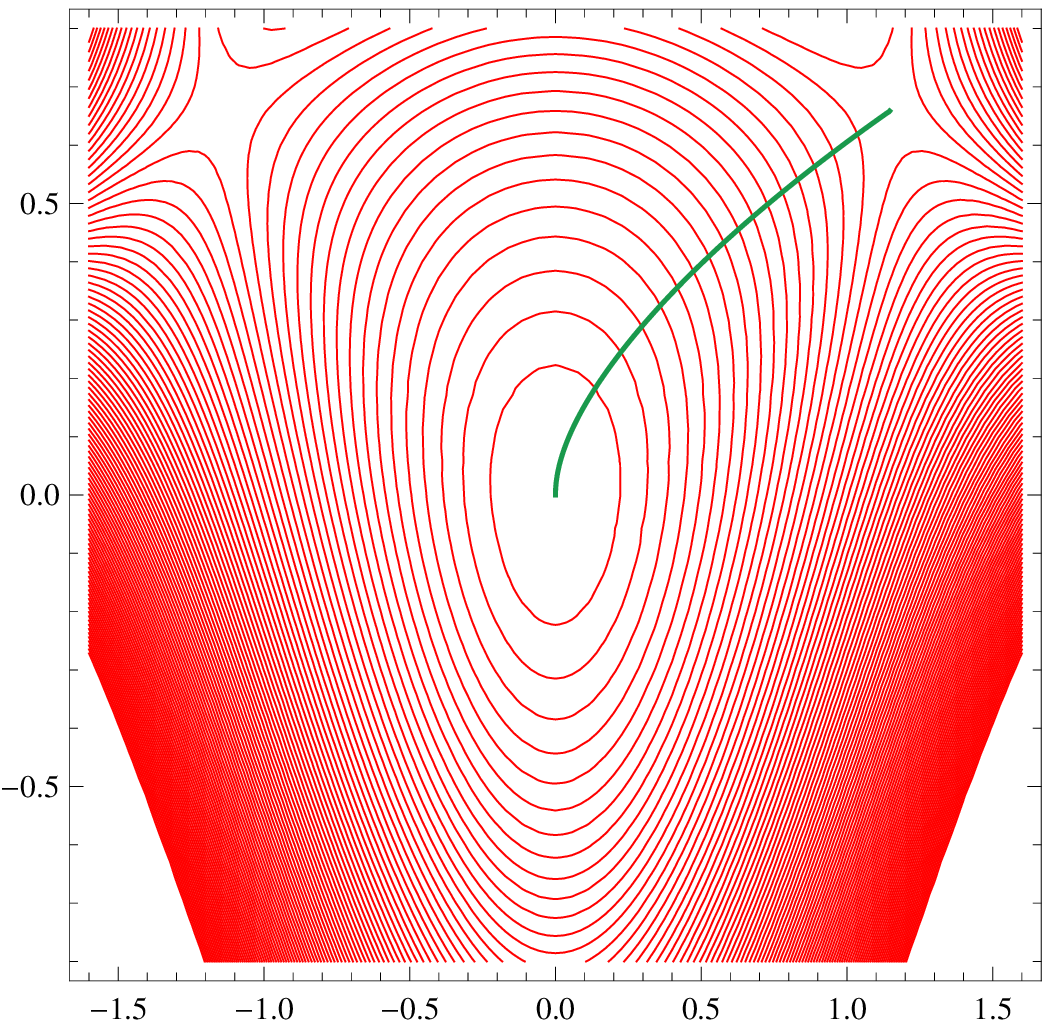}
    \caption{
     Contours of the scalar potential and
      superpotential for $\lambda_1 = \lambda_2$, $\varphi_0 =
      -\frac{\pi}{2} $ and $\varphi_1 = \varphi_2 = \frac{\pi}{4}$.
      The vertical axis is $\lambda_0$ and the horizontal axis is
      $\lambda_1 = \lambda_2$.  The central critical point is
      maximally supersymmetric.  The two critical points with
      $\lambda_0=0$ and $\lambda_1 \ne 0$ are copies of the 
      $SO(3) \times SO(3)$
      non-supersymmetric critical point.  The four other critical
      points of the potential preserve $\Neql 1$ supersymmetry, but
      the pair with $\lambda_0< 0$ preserves $\varepsilon^7
      -\varepsilon^8$ while the pair with $\lambda_0> 0$ preserves
      $\varepsilon^7 + \varepsilon^8$.  The second contour plot shows
      the superpotential that arises from preserving $\varepsilon^7 +
      \varepsilon^8$ and thus only has two non-trivial (and
      equivalent) critical points. The curving line on the plot of the
      superpotential shows the $\Neql 1$ supersymmetric flow from the
      maximally supersymmetric point to the $\Neql 1$ supersymmetric
      critical point.}
\label{Pic1}
\end{figure}

The former is the non-supersymmetric $SO(3) \times SO(3)$ invariant
critical point discovered in \cite{Warner:1983du}, with the $SO(3)$'s
acting on the coordinates $(x_3,x_6,x_7)$ and $(x_4,x_5,x_8)$.  The
second point is one of the new critical points that was recently
discovered numerically in
\cite{Fischbacher:2009cj,Fischbacher:2010ki}.  As we will see, this
point is indeed supersymmetric, as was strongly suggested by the
numerical computations.  The contours of the scalar potential for
$\lambda_1 = \lambda_2$, $\varphi_0 = -\frac{\pi}{2} $ and $\varphi_1 =
\varphi_2 = \frac{\pi}{4}$ are shown in the first graph in
Fig. \ref{Pic1}.

\subsection{The superpotential}
\label{superpot}

Define the complex superpotential by: 
\begin{equation}
\label{Wcmplx}
\begin{split}
\cW &~=~{1- \zeta_1^2\zeta_2^2 - ( \zeta_1^2-\zeta_2^2)\,\zeta_0\over (1-|\zeta_1|^2)(1-|\zeta_2|^2)(1-|\zeta_0|^2)^{1/2}} \\[6 pt]
&~=~  \cosh (\coeff{1}{\sqrt{2}} \lambda_0) \, \Big( \cosh^2 (\coeff{1}{2} \lambda_1) \cosh^2 (\coeff{1}{2} \lambda_2) -e^{-2i(\varphi_1 - \varphi_2)}\sinh^2 (\coeff{1}{2} \lambda_1) \sinh^2 (\coeff{1}{2} \lambda_2) \Big)   \\[6 pt]
 & \qquad + e^{-i \varphi_0}  \sinh (\coeff{1}{\sqrt{2}} \lambda_0) \, \Big( \cosh^2 (\coeff{1}{2} \lambda_1) \sinh^2 (\coeff{1}{2} \lambda_2)e^{2i \varphi_2} - \sinh^2 (\coeff{1}{2} \lambda_1) \cosh^2 (\coeff{1}{2} \lambda_2)e^{-2i \varphi_1}  \Big)  \,.
\end{split}
\end{equation}
Then the scalar potential is given by:  
\begin{eqnarray}
\cP &=&  4 g^2 \,\sum_{j=0}^2  \bigg( \frac{\partial  |\cW |}{\partial \lambda_j} \bigg)^2 ~+~ \frac{8 \, g^2 }{\sinh^2 (\sqrt{2}\, \lambda_0) } \bigg( \frac{\partial  |\cW| }{\partial \varphi_0} \bigg )^2 ~+~  4 \,g^2 \, \sum_{j=1}^2 \frac{1}{\sinh^2 (\lambda_j) } \bigg (\frac{\partial  |\cW |}{\partial \varphi_j} \bigg )^2   ~-~ 6\, g^2 \, |\cW|^2  \nonumber \\[6 pt]
&=&  8\, g^2\, \bigg[ (1-|\zeta_0|^2)^2 \, \bigg | \frac{\partial  |\cW |}{\partial \zeta_0} \bigg |^2  ~+~ \frac{1}{2}  \,  \sum_{j =1}^2 \,   (1-|\zeta_j|^2)^2\,  \bigg | \frac{\partial  |\cW |}{\partial \zeta_j} \bigg |^2 ~-~ 6\, |\cW|^2  \bigg]     \,.
 \label{WPreln1}
\end{eqnarray}
One can also show that:
\begin{equation}
\cP ~=~ 4\,\sum_{j=0}^2  \Big| \frac{\partial  \cW }{\partial \lambda_j} \Big|^2  ~-~ 6\, |\cW|^2\,,
 \label{WPreln2}
\end{equation}
which reflects the fact that $\cW$ is holomorphic up to a scale factor.

As we will establish below, $|\cW|$ defines a superpotential for the
truncated theory and its critical points represent supersymmetric
vacua.  One can easily verify that the critical points of $|\cW|$ are
given by $\lambda_0 =\lambda_1 =\lambda_2 =0$ (the $\Neql 8$ point)
and by:
\begin{equation} \label{newsusypt2}
\begin{split}
  \coeff{1}{\sqrt{2}} \lambda_0 ~=~  & \log \Big[ \frac{1}{\sqrt{2}} \,(\sqrt{3} + 1) \Big]  \,, \qquad  \lambda_1 ~=~  \lambda_2 =  \log \big[  \sqrt{3}  \pm  \sqrt{2}  \big]  \,, \\[6 pt]
& \varphi_0  ~=~ -\frac{\pi}{2}  \,, \ \    \varphi_1 ~=~ \varphi_2~=~ \frac{\pi}{4}  \,; \quad    \cP ~=~ - 12\, g^2  \,.
\end{split}
\end{equation}
Note that $|\cW|$ only has critical points with $\lambda_0 \ge 0$
whereas $\cP$ has symmetric sets of critical points under $\lambda_0
\to -\lambda_0$ (see Fig.~\ref{Pic1}).

For future reference, it is convenient to introduce the phase,
$\omega$, of the complex superpotential:
\begin{equation}
\cW ~=~  |\cW| \, e^{i \omega}\,.
 \label{omdefn}
\end{equation}
%

\section{Supersymmetric flows}
\label{susyflows}

We take the four-dimensional metric to have the standard holographic
flow form:
\begin{equation}
ds_{1,3}^2 ~=~ dr^2 ~+~ e^{2 A(r)}\big( \, \eta_{\mu
\nu} \, dy^\mu \, d y^\nu \big)  
\label{fourmet}
\end{equation}
and the Lagrangian of the scalars coupled to gravity is simply:
\begin{equation}
  \cL~=~  \coeff{1}{2}\, R ~-~  \cP ~+~ \cL_{kin.} \,.
 \label{Lag1}
\end{equation}

From \cite{de Wit:1982ig}, the fermion supersymmetry variations,
restricted to the scalar and gravitational sectors are:
\begin{eqnarray}
\delta \psi^i_\mu  &=&   2\, (\nabla_\mu\varepsilon^i ~+~ \coeff{1}{2}\,{{\cB_\mu}^i}_j \, \varepsilon^j) ~-~  \sqrt{2}\, g\, A_1^{ij} \gamma_\mu \varepsilon_j\,,
 \label{gravvar1}\\[6 pt]
\delta \chi^{ijk} &=&  - \cA_\mu{}^{ijk\ell}  \gamma^\mu\varepsilon_\ell  ~-~ 2\, g\, {A_{2\, \ell}}^{ijk}   \varepsilon^\ell \,,
 \label{fermvar1}
\end{eqnarray}
where we have explicitly written the scalar $SU(8)$ connection,
${{\cB_\mu}^i}_j$.

As remarked earlier, the residual supersymmetry of the $\Neql 1$ sector is:
\begin{equation}
\eta ~\equiv~  \varepsilon^7 + \varepsilon^8 \,, \qquad  \eta^* ~\equiv~ \varepsilon_7 + \varepsilon_8 \,,
 \label{ressusy}
\end{equation}
where $\eta$ is a Majorana spinor and  $\eta^*$ is its complex conjugate.  

\subsection{The supersymmetric flow equations}
\label{susyconds}

The supersymmetries along the flows may be written 
\begin{equation}
\eta ~=~   e^{\frac{1}{2} A(r)} \, e^{\frac{1}{2} i \omega} \,  \eta_0 \,,
 \label{susydep}
\end{equation}
where $A(r)$ is the metric function in (\ref{fourmet}), $\omega$ is
the phase of the complex superpotential defined in (\ref{omdefn}) and
$\eta_0$ is a constant spinor.  With this choice, the phase
dependences cancel on both sides of (\ref{fermvar1}) and one finds
that theses supersymmetry variations imply:
\begin{eqnarray}
\frac{d \lambda_j}{dr}   &=&  \pm 2 \sqrt{2}\, g \,  \frac{\partial  |\cW |}{\partial \lambda_j} \,, \quad  j=0,1,2 \,; \qquad  \frac{d \varphi_0}{dr}   ~=~ \pm  \frac{  4 \sqrt{2}\, g }{\sinh^2 (\sqrt{2} \lambda_0)} \frac{\partial  |\cW |}{\partial  \varphi_0} \,,  \nonumber \\
 \frac{d \varphi_j}{dr}     &=&   \pm \frac{  2 \sqrt{2}\, g }{\sinh^2 ( \lambda_j)} \frac{\partial  |\cW |}{\partial  \varphi_j} \,, \quad  j= 1,2 \,,
   \label{floweqn1}
\end{eqnarray}
where the signs are determined by the choice of the radial helicity
projector on the supersymmetry:
\begin{equation}
\gamma^r \, \eta_0 ~=~   \pm  \eta_0 \,.
 \label{helproj}
\end{equation}

As usual, the complex superpotential is defined using the eigenvalues
of the $A_1$-tensor:
\begin{equation}
A_1^{ij} \, v^j ~=~   \cW \, v^j  \,, \qquad \vec v ~\equiv~ (0,0,0,0,0,0,1,1)  \,.
 \label{Wdefn}
\end{equation}
The gravitino variations in the $y^\mu$ directions of (\ref{fourmet})
then give the usual flow equation:
\begin{equation}
 \frac{d A(r)}{dr}    ~=~   \mp \sqrt{2}\, g\,   |\cW | \,.
   \label{floweqn2}
\end{equation}

The last supersymmetry condition is the radial component of the
gravitino variation and this requires the non-trivial identity coming
from the cancellation of the derivatives of the phase, $\omega$, and
terms arising from the $SU(8)$ connection, ${{\cB_\mu}^i}_j$:
\begin{equation}
\frac{i}{2}\,  \frac{d \omega}{dr}   ~=~    i\, \sinh^2 (\coeff{1}{\sqrt{2}} \,\lambda_0) \, \frac{d \varphi_0}{dr} ~+~ 2\, i\,  \sinh (\coeff{1}{2}  \, \lambda_1)\, \frac{d \varphi_1}{dr}  ~-~ 2\,  i\,  \sinh ( \coeff{1}{2}  \,\lambda_2)  \,  \frac{d \varphi_2}{dr} \,.
   \label{omconsistent}
\end{equation}
One can indeed verify that this equation is satisfied by virtue of the
definition of $\omega$ in (\ref{omdefn}) and the flow equations
(\ref{floweqn1}).

Thus (\ref{floweqn1}) and (\ref{floweqn2}) generate families of
supersymmetric flows in the $\Neql 1$ theory defined 
 in Section \ref{contrunc}.  In particular,
critical points of $|\cW|$, like that given in (\ref{newsusypt2}),
define supersymmetric vacua.

\subsection{The new supersymmetric flow}
\label{susyflow}

We will choose the lower  sign in (\ref{floweqn1}) and
(\ref{floweqn2}) so that $A'(r) >0$. One then finds that
(\ref{floweqn1}) implies
\begin{equation}
 \frac{d |\cW|}{dr}   ~<~   0 \,.
   \label{monotonic1}
\end{equation}
along all flows away from critical points.  Thus, as the $c$-theorem
requires, $A'(r)$ and $|\cW|$ increase as $r$ decreases from infinity.
One can easily check that
\begin{equation}
 \frac{d |\cW|}{d \varphi_0}   ~=~   \frac{d |\cW|}{d \varphi_1}   ~=~  \frac{d |\cW|}{d \varphi_2}   ~=~  0 \quad {\rm for} \quad \varphi_0 = -  \frac{\pi}{2}\,, \ \  \varphi_1 = \varphi_2 =   \frac{\pi}{4}\,,     \label{fixangs1}
\end{equation}
but with {\it arbitrary} $\lambda_j$.  One can also check that 
\begin{equation}
 \frac{d |\cW|}{d \lambda_1}~-~ \frac{d |\cW|}{d \lambda_2}     ~=~   0  \quad {\rm for} \quad  \lambda_1 =     \lambda_2\,, \ \    \varphi_0 = - \frac{\pi}{2}\,, \ \  \varphi_1 =    \varphi_2 \,.    \label{fixlamdas1}
\end{equation}
In particular, it is consistent with the flow equations
(\ref{floweqn1}) to restrict to the space $ \lambda_1 = \lambda_2$, $
\varphi_0 = -\frac{\pi}{2}$, $\varphi_1 = \varphi_2 = \frac{\pi}{4}$
upon which one has a superpotential:
\begin{equation}
\cW  ~=~ \cosh ( \lambda_1)\, \cosh (\coeff{1}{\sqrt{2}} \,\lambda_0)\,~-~  \frac{1}{2}\,\sinh^2 ( \lambda_1)\,   \sinh (\coeff{1}{\sqrt{2}} \,\lambda_0)  \,.    \label{restsuper}
\end{equation}
This superpotential is plotted in the second graph in Fig.~\ref{Pic1}.
One can easily see that there is a steepest descent solution to the
flow between the $\Neql 8$ critical point and the new $\Neql 1$
critical point.  We have solved this flow numerically and the result
is plotted in the second graph in Fig.~\ref{Pic1}.

In the neighborhood of the $\Neql 8$ critical point the superpotential
in (\ref{restsuper}) has the expansion
\begin{equation}
\cW  ~=~ 1 ~+~  \coeff{1}{4}\, \lambda_0^2~+~  \coeff{1}{2}\, \lambda_1^2 ~-~ \coeff{1}{\sqrt{2}}\, \lambda_0\,  \lambda_1^2 ~+~ \dots \,.    \label{supersmall}
\end{equation}
The flow equations  are simply:
\begin{equation}
\frac{d \lambda_0}{dr}  =  -  \frac {2}{L} \,  \frac{\partial \cW }{\partial \lambda_0} \approx - \frac {1}{L} \, \lambda_0\,,   \qquad
\frac{d \lambda_1}{dr}  =  - \frac {2}{L} \,  \frac{\partial \cW }{\partial \lambda_1} \approx -  \frac {2}{L}\, \lambda_1  \,, \qquad  A'(r) =  \frac{1}{L} \cW \approx \frac{1}{L}  \,,
   \label{floweqn3}
\end{equation}
where we have set $g = \frac{1}{\sqrt{2}L}$.  Hence the flow starts out with:
\begin{equation}
\lambda_0  ~\approx~ a_0 \, e^{-{r/L}} \,,   \qquad  \lambda_1  ~\approx~ a_1 \, e^{-2{r/L}} \,,    
   \qquad r~\rightarrow~\infty\,, \label{asymp1}
\end{equation}
for some constants $a_0$ and $a_1$.  This explains the nearly
parabolic behavior of the flow near the origin in Fig. \ref{Pic1}.

It is very interesting to note that $\lambda_0$ flows out along a {\it
  non-normalizable} mode in $AdS_4$ while $\lambda_1$ flows out along
a {\it normalizable} mode in $AdS_4$.  Also note that unlike the
massive flows considered in \cite{Freedman:1999gp}, the leading UV
behavior for $\lambda_0$ (parametrized by $a_0$) does not source the
leading behavior in $\lambda_1$.  These flows are thus independent in
the UV except for the fine tuning required by targeting the new fixed
point.

\subsection{The mass matrices at the new critical point.}
\label{masses1}

\begin{table}[t]
\begin{center}
\scalebox{0.8}{
\begin{tabular}{@{\extracolsep{25 pt}}c c c c }
\toprule
\noalign{\smallskip}
$\#$ of modes \hspace{-10 pt} & $m^2L^2$ & $SO(2)\times SO(2)$ irrep & $\cN=1$ scalars\\
\noalign{\smallskip}
\midrule
\noalign{\smallskip}
4 & $2+\sqrt{15}+\sqrt{4+\sqrt {15}} ~\approx~ 8.679$ & $(2,2)$ & \\[10 pt]
1 & $3(1+\sqrt 3)~\approx~ 8.196$ & $(1,1)$  & 1\\[10 pt]
4 & $\displaystyle {5\over 2}+\sqrt {10}~\approx ~ 4.412 $ & $(2,1)\oplus(1,2) $ & \\[10 pt]
4 & $2+\sqrt{15}-\sqrt{4+\sqrt {15}} ~\approx~ 3.067$  &  $(2,2)$ & \\[10 pt]
1 & $1+\sqrt 3 ~\approx ~ 2.732 $ &  $(1,1)$ & 1\\[10 pt]
& & $(2,2)_{\times 2}$  &   \\
30 & $0$ & $ (2,1)_{\times 4}\oplus(1,2)_{\times 4}$  & \\
& & $  (1,1)_{\times 6} $ &   \\[10 pt]
1 & $1-\sqrt 3~\approx~ -0.732 $ &  $(1,1)$ & 1 \\[6pt]
4 & $\displaystyle -{5\over 4}\eql -1.25  $ &  $(2,1)\oplus(1,2) $  & \\[6pt]
4 & $ \displaystyle 2- \sqrt{{ 3\over 2}} + \sqrt{{5\over 2}}-\sqrt{15} ~\approx ~ -1.517 $ &  $(2,2)$  & \\[10 pt]
4 & $\displaystyle {5\over 4}-\sqrt{10}~\approx~ -1.912 $ &  $(2,1)\oplus(1,2) $  & \\[10 pt]
4 & $-2$ & $4\times (1,1)$ & 2 \\[10 pt]
1 & $3(1-\sqrt 3 ) ~\approx~ -2.196 $ & $(1,1)$  & 1 \\[10 pt]
4 & $2-\sqrt{15}-\sqrt{4+\sqrt {15}} ~\approx~ -2.229 $ &  $(2,2)$ & \\[10 pt]
4 & $\displaystyle -{9\over 4}\eql -2.25 $ &  $(2,1)\oplus(1,2) $  & \\[10 pt]
\bottomrule
\end{tabular}
}
\caption{
\label{tblsusy}
 The  spectrum of scalars at the new $\Neql 1$ supersymmetric critical point.  }
\end{center}
\end{table}

The mass matrix for a general scalar fluctuation can be found in
\cite{deWit:1983gs}.  For a general scalar fluctuation,
$\Sigma_{ijkl}$, one has, at quadratic order,
\begin{equation}\label{scalmassmat}
\begin{split}
 \cL(\Sigma^2)  & = - \coeff{1}{96}\,  \,g^{\mu \nu} \partial_\mu  \Sigma_{ijkl} \, \partial_\nu\Sigma^{ijkl}  - \coeff{g^2}{96}\, \Big( \big(\coeff{2}{3}\, \cP + \coeff{13}{72}\, \big| {{A_2}_\ell}^{ijk} \big|^2 \big) \, \Sigma_{ijkl} \,  \Sigma^{ijkl}   \\[6 pt] 
& \qquad ~+~  \big(6\,  {{A_2}_k}^{mni}   {{A_2}^j}_{mn\ell} ~-~ \coeff{3}{2}\, {{A_2}_n}^{mij}   {{A_2}^{n}}_{mk\ell}  \big) \, \Sigma_{ijpq} \,  \Sigma^{klpq}   \\[6 pt] 
& \qquad ~-~ \coeff{2}{3}\, {{A_2}^i}_{mnp}   {{A_2}_{q}}^{jk\ell}    \, \Sigma^{mnpq} \,  \Sigma_{ijkl}  \Big) \,.   
\end{split}
\end{equation}
If one uses this formula at the $\Neql 1$ critical point one finds the
scalar spectrum given in Table~\ref{tblsusy}.  
The first column in the table contains the degeneracies, the second column  has the masses and the third shows the  charges under the residual symmetry.  The last column shows the location of six scalars that are in the  $\Neql 1$ theory analyzed in Section \ref{Neql1sector}. Those scalars are   thus singlets under the  symmetry that defined the truncation.  

Note that all the masses obey the BF bound, as they must. 
It is interesting to observe that there are four fields that saturate the
BF bound and there are four other fields whose dimension differs by
one unit form these BF saturating fields.  It seems likely that these
fields form an $\Neql 1$ supermultiplet.  It would also be interesting to 
see if these fields might be used for some form of  Coulomb branch flow 
much like that investigated in \cite{Khavaev:2000gb,Khavaev:2001yg}.  
These fields are charged under the two residual $U(1)$'s and so such
a flow would break all  the symmetry.

\section{The flow in the holographic field theory }
\label{holFT}

In the neighborhood of the maximally supersymmetric point, the seventy
scalars of the gauged supergravity theory are holographically dual to
the (traceless) bilinears in the scalars and fermions:
\begin{eqnarray}
 \cO^{IJ}  &=&  \Tr~\big(X^I \, X^J) ~-~ \coeff{1}{8}\, \delta^{IJ}\, \Tr~\big(X^K \, X^K\big) \,, \quad  I,J, \dots =1,\dots,8 \\
  \cP^{AB} &=&  \Tr~\big(\psi^A \, \psi^B\big) ~-~ \coeff{1}{8}\, \delta^{AB}\, \Tr~\big(\psi^C \, \psi^C \big) \,, \quad A,B, \dots =1,\dots,8\,, 
\end{eqnarray}
where $\cO^{IJ}$ transforms in the ${\bf 35}_s$ of $SO(8)$, and
$\cP^{AB}$ transforms in the ${\bf 35}_c$.  The real parts of $w_j$ in
(\ref{Liescalars}) can thus be thought of as the duals of some of the
$\cO^{IJ}$ and the imaginary parts of $w_j$ can be thought of as the
duals of some of the $\cP^{IJ}$.

To make this map more specific, if one sets all of the angles
$\varphi_j = 0$, then the scalar fields of the $\Neql 1$ theory all
lie in the $SL(8, \IR)$ subgroup of $E_{7(7)}$ and if one maps to this
basis then one finds that the corresponding $SL(8, \IR)$ matrix may be
written:
\begin{equation}   \label{SL8mat}
\cS  ~=~  {\rm   diag} \big(e^{\mu_1}\,,  e^{\mu_1}\,, e^{\mu_2}\,, e^{\mu_2}\,, 
e^{\mu_3}\,, e^{\mu_3}\,, e^{\mu_4}\,, e^{\mu_4} \big)  \,, 
\end{equation}
\begin{equation}
\begin{split}
\mu_1 &~\equiv~ \frac{1}{2 }\, \lambda_1 -  \frac{1}{2\sqrt{2} }\,\lambda_0\,, \qquad  \mu_2 ~\equiv~ -  \frac{1}{2 }\, \lambda_1 -  \frac{1}{2\sqrt{2} }\,\lambda_0\,, \\[6 pt]
\mu_3 &~\equiv~  \frac{1}{2 }\, \lambda_2 +  \frac{1}{2\sqrt{2} }\,\lambda_0\,, \qquad \mu_4 ~\equiv~  - \frac{1}{2 }\, \lambda_2 + \frac{1}{2\sqrt{2} }\,\lambda_0\,.
\end{split}
\end{equation}
The fact that the critical flow has $\varphi_0 = - {\pi \over 2}$ and
$\varphi_1 = \varphi_2 = {\pi \over 4}$ means that the scalar field
represented by $w_0$ lies entirely within the ${\bf 35}_c$ sector
while $w_1$ and $w_2$ are equally split between the two sectors.
Moreover, we have $w_1 = w_2$ and so the operators that are involved
in the $\Neql 1$ flow described above may be written:
\begin{eqnarray}
 \cO^{IJ}    &=&   {\rm   diag} \big(e^{\nu_1}\,,  e^{\nu_1}\,, e^{\nu_1}\,, e^{\nu_1}\,, e^{-\nu_1}\,, e^{-\nu_1}\,, e^{-\nu_1}\,, e^{-\nu_1} \big)  \,,    \\[6 pt]
  \cP^{AB}   &=&   {\rm   diag}\big(e^{\nu_0 + \nu_1}\,,  e^{\nu_0 + \nu_1}\,, e^{\nu_0 - \nu_1}\,, e^{\nu_0 - \nu_1}\,, e^{-\nu_0 + \nu_1}\,, e^{-\nu_0+ \nu_1}\,, e^{-\nu_0 - \nu_1}\,, e^{-\nu_0 - \nu_1} \big)  \,,   \\[6 pt]
\nu_0&\equiv&   \frac{1}{2\sqrt{2} }\, \lambda_0 \,,   \qquad  \nu_1~\equiv~  \frac{1}{2\sqrt{2} }\, \lambda_1 ~=~  \frac{1}{2 \sqrt{2}}\, \lambda_2 \,.
   \label{holOps}
\end{eqnarray}

Non-normalizability of the $\lambda_0$ mode suggests that $\nu_0$ is
dual to a mass term while normalizability of the $\lambda_1$ mode
suggests that $\nu_1$ might be dual to a vacuum expectation value.
Thus we appear to have a flow that involves an $SO(4) \times
SO(4)$-invariant fermion mass term combined with some boson vevs and
fermion condensates.  If one sets $\nu_1=0$ and looks at the pure
fermion mass flows then these preserve the sixteen supersymmetries of
the $\Neql 4$ supergravity and indeed were studied extensively in
\cite{Pope:2003jp, Bena:2004jw,Lin:2004nb}.  In particular, such flows
have infra-red fixed points that may be described, at least in the infra-red,  
in terms of free fermions in $(1+1)$-dimensions \cite{Lin:2004nb} and whose
perturbations describe excitations of the Fermi sea.  The new
supersymmetric critical point involves adding some {\it normalizable
  fluxes} to this picture, which means that the flow involves a
modification of the {\it state} of the holographic field theory and
not a change of Lagrangian.  Therefore, our new $\Neql 1$ flow has a
potentially very interesting interpretation as a new (stable,
supersymmetric) state of the fermion droplet theory and this state
involves some form of bosonic vev and Fermi condensate.

It is also interesting to note that for this flow, the holographic
central charge changes in a very simple way:
\begin{equation}
 {c_{\rm IR} \over c_{\rm UV}} ~=~  \left( {\cP_{\rm IR} \over {\cP_{\rm UV}} }\right)^{-1} ~=~ {1 \over 2}  \,.
   \label{cratio}
\end{equation}
%
 
\section{Stability of the  $SO(3) \times SO(3)$-invariant point}
\label{masses2}

\begin{table}[t]
\begin{center}
\scalebox{0.8}{
\begin{tabular}{@{\extracolsep{25 pt}}c c c c }
\toprule
\noalign{\smallskip}
$\#$ of modes \hspace{-10 pt} & $m^2L^2$ & $SO(3)\times SO(3)$ irrep & $\cN=1$ scalars\\
\noalign{\smallskip}
\midrule
\noalign{\medskip}
1 & $\displaystyle {60\over 7}~\approx~ 8.571$ & $(1,1)$ &  1  \\[10 pt]
9 & $\displaystyle {3\over 7} \, (\,5+\sqrt{65}\, )~\approx~5.598 $ & $(3,3)$ & 1 \\[10pt]
9 & $\displaystyle {18\over 7}~\approx~2.571$ & $(3,3)$ & 1 \\[10pt]
22 & $0$ & $(3,3)\oplus (3,1)_{\times 2}\oplus(1,3)_{\times 2}\oplus (1,1)$ & \\[10pt]
9 & $\displaystyle {3\over 7} \, (\,5-\sqrt{65}\, )~\approx~-1.312$ & $(3,3)$ & 1 \\[10pt]
20 & $\displaystyle -{12\over 7} ~\approx~ -1.714 $ & $(3,3)_{\times 2}\oplus(1,1)_{\times 2}$ & 2 \\[10pt]
\bottomrule
\end{tabular}
}
\caption{\label{tbloldso3}
The  spectrum of scalars at the old $SO(3) \times SO(3)$-invariant critical point.  
}
\end{center}
\end{table}

If one uses (\ref{scalmassmat}) at the critical point with
residual $SO(3)\times SO(3)$ symmetry and $\cP=-14\,g^2$,
one finds the scalar spectrum given in Table~\ref{tbloldso3}. 
Note that all the fields obey the BF bound and so this critical point
is perturbatively stable.

The flow to this critical point cannot, of course, be supersymmetric,
but it has already been discussed in \cite{Distler:1998gb}.  Given the
perturbative stability of this critical point, it would be well worth
revisiting this flow to understand its role within $AdS/CMT$.

It is also important to note that the cosmological constant of this
critical point, $-14 g^2$, is lower than that of the new $\Neql 1$
point and, as is evident from the contours of the potential in
Fig.~\ref{Pic1}, there must be a flow from the $\Neql 1$ point
to the $SO(3) \times SO(3)$-invariant point.  The relevant operator that
drives this flow is dual to a combination of $\lambda_0$ and
$\lambda_1$ and in the neighborhood of the $\Neql 1$ point this has
$m^2 L^2 = 3(1-\sqrt{3}) \approx -2.19615$.  It would be most
interesting to understand the role of this phase in the holographic
field theory.  Since the cosmological constants for the $SO(3) \times
SO(3)$-point and the $\Neql 1$ point are integers, the ratios of
central charges at the ends of all these flows are simple rational
numbers, like ${6 \over 7}$.

\section{A  new critical point with $\sotsotgp$ invariance}
\label{newso3point}

The supersymmetric truncation in Section~\ref{Neql1sector} was found
initially using numerical results both for the continuous symmetry and
exact location of the new $\cN=1$ supersymmetric critical point (point
\#11 in \cite{Fischbacher:2009cj} and Appendix~\ref{nummasses}). We
will now discuss an example where numerical constraints on the
continuous symmetry   and the mass spectrum are sufficient
to set up a feasible analytic calculation of a new critical point.

A preliminary numerical search for   critical points beyond the   ones found in \cite{Fischbacher:2009cj} (and listed in Appendix~\ref{nummasses}) has  identified a new critical point with a relatively large symmetry group.  Indeed, numerical data for this point  given in Appendix~\ref{extrapoints} indicate six continous symmetries and a completely degenerate gravitino mass spectrum. This suggests that the symmetry group might be simply  another $SO(3)\times
SO(3)$, that is obtained by a triality rotation of the   $SO(3)\times SO(3)$  in
Section~\ref{masses2}. To distinguish between the two we will denote this new symmetry by $\sotsotgp$. It is explicitly realized by the following generators of $SO(8)$:
\begin{equation}\label{symmso3}
\begin{split}
\alpha_iT_i^{(1)}+\beta_iT_i^{(2)}   & = \alpha_1 \tau201+\alpha_2\tau002+\alpha_3\tau203\\
&\quad + \beta_1\tau102+\beta_2\tau200+\beta_3\tau302\,,
\end{split}
\end{equation}
where $\sigma^0$ is the unit matrix and $\sigma^i$, $i=1,2,3$, are the Pauli matrices. Under these generators, the supersymmetries and the eight gravitini transform in two copies of the $4$ of  $\sotsotgp$.

The eight components of a vector in the tensor product \eqref{symmso3} are ordered as $(111), (112), (121),$ etc., and correspond to the eight vector components of $SO(8)$.  There are six  invariant, noncompact generators of $E_{7(7)}$  spanned by the following forms:
\begin{equation}\label{}
\begin{split}
\Psi_0\eql w_0\,\Big[\,& \dx1234+\dx5678\\ &-\dx1357-\dx2468 \\ & -\dx1368-\dx2457\,\Big]\,,
\end{split}
\end{equation}
\begin{equation}\label{psione}
\begin{split}
\Psi_1\eql w_1\,\Big[\,& \dx1278+\dx3456\\ &+\dx1458+\dx2367 \\ & -\dx1467-\dx2358\,\Big]\,,
\end{split}
\end{equation}
\begin{equation}
\begin{split}
\Psi_2\eql   w_2\,\Big[\,& \dx1258-\dx1267\\ &+\dx1456-\dx2356 \,\Big]\\
-\overline w_2\,\Big[\,& \dx1478-\dx2378\\ &+\dx3458-\dx3467 \,\Big]\,,
\end{split}
\end{equation}
\begin{equation}\label{psithree}
\begin{split}
\Psi_3\eql w_3\,\dx1256+\overline w_3\,\dx3478\,,
\end{split}
\end{equation}
where 
\begin{equation}\label{}
w_0\eql {1\over\sqrt 3}\lambda_0\,,\qquad w_1\eql  {1\over\sqrt 3}\lambda_1\,,\qquad w_2\eql -{1\over 2}(\lambda_2-i\lambda_4)\,,\qquad w_3\eql \lambda_3+i\lambda_5\,.
\end{equation}
The real parameters $\lambda_i$, $i=0,\ldots,5$, correspond to canonically normalized generators and parametrize the coset space:
\begin{equation}\label{}
O(1,1)\times {SL(3,\RR)\over SO(3)}\,,
\end{equation}
where $\lambda_0$ is the coordinate on the first factor, while $\lambda_1,\ldots,\lambda_5$ are coordinates on the second factor with the corresponding noncompact generators of $SL(3,\RR)$  given by
\begin{equation}\label{}
\Lambda\eql \left(\begin{matrix}
\frac{2 }{\sqrt{3}}\lambda _1 & \lambda _2 & \lambda _4 \\
 \lambda _2 & -\frac{1}{\sqrt{3}}\lambda _1 +\lambda _3& \lambda _5 \\
 \lambda _4 & \lambda _5 & -\frac{1}{\sqrt{3}}\lambda _1-\lambda _3
\end{matrix}\right)\,.
\end{equation}

The unbroken $O(2)$ gauge symmetry acts on $\Lambda$ as the $\cR_{12}$
rotation and allows one to set a linear combination of $\lambda_4$ and
$\lambda_5$ to zero. This reduces the number of independent parameters
in the potential to five.

To calculate the potential, one must exponentiate the $E_{7(7)}$
generators $\Psi_0$ and $\Psi=\Psi_1+\Psi_2+\Psi_3$. While this is
completely straightforward for the first generator, it is extremely
difficult for the second one if one insists on keeping
$\lambda_1\,,\ldots,\lambda _5$ as explicit parameters. Instead, we
express the potential as a function of the matrix elements, $m_{ij}$,
of the $SL(3,\RR)$ group element
\begin{equation}\label{expM}
M\eql e^{\Lambda}\,,\qquad M\eql (m_{ij})_{i,j=1,2,3}\,.
\end{equation}
Note that by construction $M$ is symmetric, $m_{ij}=m_{ji}$. The exponentiation is accomplished by   a similarity transformation, $S$, that brings $\Psi$ to the block diagonal form,
\begin{equation}\label{}
S^{-1}\Psi S\eql {\rm diag}(6\times (\Lambda,-\Lambda),20\times 0)\,.
\end{equation}

The potential is a sixth-order polynomial in $\rho=e^{-2\lambda_0/\sqrt 3}$ 
and the $m_{ij}$'s. It
becomes algebraically quite manageable if we use the remaining gauge
symmetry to set a linear combination of the matrix elements $m_{13}$
and $m_{23}$ to zero. It can then be further simplified by solving
explicitly the unimodularity condition, $\det M\eql 1$, to eliminate
one additional matrix element. The final result is given in
Appendix~\ref{newpotso3}.

It is clear that a systematic search for critical points of the full
potential \eqref{so3pot} is still quite involved, and we have not
carried it out in detail. In the following we consider two natural
restrictions in which we keep only two commuting fields: the $O(1,1)$
field, $\lambda_0$, and one additional field, $\lambda$, in
$SL(3,\RR)$.

First,  take $\Lambda$, and hence $M$,  to be diagonal by setting
\begin{equation}\label{so7tr}
\lambda_1\eql {\sqrt 3\over 2}\,\lambda\,,\qquad \lambda_3\eql -{1\over 2}\,\lambda\,,\qquad \lambda_2\eql\lambda_4\eql\lambda_5\eql 0\,.
\end{equation}
The restriction of the potential \eqref{so3pot} to $\lambda_0$ and $\lambda$ is
\begin{equation}\label{so7pot}
\cP\eql -g^2\,\left[\,3\,\rho^{-1}+3\,\rho\,\cosh(2\lambda)+{1\over 4}\,\rho^3 (1-\cosh(4\lambda))\,\right] \,,\qquad \rho\eql e^{-2\lambda_0/\sqrt 3}\,,
\end{equation}
and, apart from the trivial point $\lambda_0=\lambda=0$, there is one nontrivial critical point at
\begin{equation}\label{so7pt}
\lambda_0\eql -{\sqrt 3\over 8}\,\log (5)\,, \qquad \lambda\eql \pm \,{1\over 4}\,\log(5)\,,\qquad \cP\eql -2 \cdot  5^{3/4}\,g^2\,.
\end{equation}

The value of the cosmological constant identifies this point as the
$SO(7)^+$ critical point. By expanding \eqref{so7pot} to the quadratic
order and using that $\lambda_0$ and $\lambda$ have canonical kinetic
terms, we find two masses: $m^2L^2= 6$ and $ -{12/ 5}$. Those are
indeed correct values for the masses of scalar fluctuations at this
point \cite{deWit:1983gs} (see, also \cite{Bobev:2010ib} and
Table~\ref{tbl:scalarmasses}), which suggests that \eqref{so7tr} is a
consistent truncation in this sector. Furthermore, we see that there
is one unstable mode, which arises as a linear combination of the two
modes in \eqref{so7tr}.

\begin{table}[t]
\begin{center}
\scalebox{0.8}{
\begin{tabular}{@{\extracolsep{25 pt}}r c c   }
\toprule
\noalign{\smallskip}
$\#$ of modes \hspace{-10 pt} & $m^2L^2$ & $\sotsotgp\simeq SO(3)\times SO(3)$ irrep \\
\noalign{\smallskip}
\midrule
\noalign{\medskip}
1 & $3+\sqrt{3}+\sqrt{6 (4+\sqrt{3})}~\approx~ 10.597$ &   $(1,1)$  \\[10 pt]
1 & $\displaystyle {3\over 2} (5+\sqrt{3})~\approx~    10.098$ & $(1,1)$ \\[10 pt]
1 &  $ 4 \sqrt{3} ~\approx~ 6.928 $ &    $(1,1)$  \\[10 pt]
9 & $2 (\sqrt{3}-1)~\approx~ 1.464$  &  $(3,3)$   \\[10 pt]
22 &  $0$ &  $(3,3)\oplus(3,1)_{\times 2}\oplus (1,3)_{\times 2}\oplus (1,1)$   \\[10 pt]
9 & $2 (\sqrt{3}-2)~\approx~ -0.536$ &   $(3,3)$  \\[10 pt]
1 & $3+\sqrt{3}-\sqrt{6 (4+\sqrt{3})}~\approx~ -1.132$ &  $(1,1)$   \\[10 pt]
15 & $\displaystyle\frac{3}{2} (\sqrt{3}-3)~\approx~ -1.902$ &    $(3,3)\oplus(3,1) \oplus (1,3)  $  \\[10 pt]
10 & $2 (\sqrt{3}-3)~\approx~ -2.536$ &    $(5,1)\oplus(1,5)$ \\[10 pt]
1 & $-2 \sqrt{3}~\approx~ -3.464$ &     $(1,1)$ \\[10 pt]
\bottomrule
\end{tabular}
}
\caption{\label{tbl:newso3} The spectrum of scalars at the $\sotsotgp$ point.  }
\end{center}
\end{table}

The second natural restriction is with completely off-diagonal
$\Lambda$. It is clear from the form of the generators $\Psi_i$ in
\eqref{psione}-\eqref{psithree} that the simplest choice is to take

\begin{equation}\label{newtr}
\lambda_1\eql\lambda_2\eql\lambda_3\eql\lambda_4\eql0\,,\qquad \lambda_5\eql \lambda\,.
\end{equation}
This leads to the potential 
\begin{equation}\label{newpot}
\cP\eql -{g^2\over 4}\,\rho^{-1} \left[\, (\rho^4-6\rho^2-3)\,\cosh(2\lambda)-(\rho^2+3)^2\,\right]\,,\qquad \rho\eql e^{-2\lambda_0/\sqrt 3}\,.
\end{equation}
A straightforward calculation reveals one nontrivial critical point at
\begin{equation}\label{}
\lambda_0\eql -{\sqrt 3\over 4}\,\log (3+2\sqrt 3 )\,,\qquad \lambda\eql \pm {1\over 2}\,{\rm arccosh}(\sqrt 3)\,,
\end{equation}
with the cosmological constant
\begin{equation}\label{}
\qquad \cP\eql -6\,\sqrt{1+{2\over \sqrt 3}}\,g^2~\approx~ -8.807\,g^2\,.
\end{equation}
Indeed, this is the same value as found in the numerical search in Appendix~\ref{extrapoints}. The calculation of
the scalar masses is summarized in
Table~\ref{tbl:newso3} and agrees with the numerical result. The two masses on the restricted
fields are $m^2L^2= 3\sqrt 3$ and $-2\sqrt 3\approx -3.464$, which
once more suggests a consistent truncation while the latter exhibits one of the unstable
modes at this point. Other unstable modes transform in $(5,1)\oplus(1,5)$ of $SO(4)'$ and hence are not visible in this truncation.

\section{The new critical points with at most $U(1)^2$ symmetry}
\label{OtherPts}

Given that the $SO(8)$ gauged $\mathcal{N}=8$ model indeed {\em does}
have non-supersymmetric perturbatively stable vacua, an obvious
question is how many there are and, given that the total number of
critical points may well be very large, whether or not stability without
supersymmetry is a rare.

Using the numerical data provided in~\cite{Fischbacher:2010ki} to
study the stability of the fourteen new critical points that were
presented in~\cite{Fischbacher:2009cj} to determine scalar masses, one
finds that the thirteen solutions without residual supersymmetry all
violate the BF bound $m_{\rm scalar}^2L^2\ge -9/4=-2.25$ so strongly that this 
cannot be attributed to the limited numerical accuracy to which their
positions have been determined. Details on the scalar mass matrix
eigenvalues and their degeneracies are given in
Appendix~\ref{nummasses}.  This seems to suggest that stability
without supersymmetry indeed is a rare phenomenon in
four dimensions.\footnote{In three dimensions  the situation is almost the opposite:  
One finds that most of the three-dimensional, non-supersymmetric critical points are BF-stable.} 
It is almost a comical coincidence that the very first critical point that was found
through a systematic analysis of the scalar
potential~\cite{Warner:1983du} is, at present, the only known one that is
non-supersymmetric and stable!

\subsection{Genericity of the new critical points}

Another important question about all these new critical points is how 
generic they are and to what  extent to which they represent an 
exhaustive list or whether they are a ``random sample'' of perhaps 
many undiscovered points.  

There are two important aspects to this issue. First, the numerical method uses
the  sensitivity back-propagation method to minimize $|Q|^2$, where $Q_{ijkl}$ is
the tensor self-duality condition that defines a critical point  
 \cite{deWit:1983gs, Fischbacher:2009cj}.  The algorithm starts from a choice 
of a random  point but even so, it provides a remarkably
efficient strategy to numerically obtain a lot of useful information
about previously unknown stationary points which may then be utilized
as an input to a fully analytic investigation.   However, the size of
the basin of attraction seems to vary strongly between different
stationary points. Numerical searches can have four different
outcomes: (a) the calculation fails due to a numerical overflow, (b)
the calculation `gets stuck' as optimization proceeds extremely
slowly, (c) the calculation produces a stationary point which already
was found earlier, and (d) the calculation produces a novel
solution. With some code tweaking, (a) can be avoided almost
completely, (b) happens rarely, but quite typically in the
neighborhood of some `tricky' critical points (the $\Neql 1$ vacuum
with $\cP/g^2=-12$ being one of them), (c) is fairly frequent but with
a very uneven distribution of results, and (d) happens sufficiently
often to assume that the total number of solutions is far larger than
those described so far. 

The second aspect is that typical violations
of the stationarity condition, measured by $|Q|^2$, get larger the
further one moves outward from the $\Neql 8$ vacuum on the scalar
manifold. Hence, gradients naturally tend to draw the numerical
optimization towards the $\Neql 8$ point. In order to counteract this,
the numerical search strategy which was developed
in~\cite{Fischbacher:2008zu} for $\Neql 16$ theory in three dimensions and also employed
to find the fourteen new solutions comes with a `tweaking parameter' that
essentially allows some control over the distance from the $\Neql 8$
point at which optimization will tend to spend most
time. In~\cite{Fischbacher:2009cj}, this parameter was chosen to scan
at a distance somewhat beyond the $SU(3)$-invariant critical
points. Due to the existence of this tweaking parameter, all claims
about how generic the solutions thus found by this approach must be
qualified with a statement about the search range.

While a far more detailed analysis of the scalar potentials of all
(highly extended) supergravity models is now possible compared to what
would have been considered as feasible some years ago, there also are
some indications (considering results for the $\Neql 16$ theory in
three dimensions \cite{Fischbacher:2008zu}) that hardware-supported
IEEE-754 floating point numbers might be insufficient to find and
analyze solutions that lie `very far out'. Apart from this,
high-precision numerics also is important for semi-automatically
producing analytic conjectures from numerical data, as explained
in~\cite{Fischbacher:2010ki}. This will be explored in more detail in
the next sub-sections.

One generally would expect that analytic expressions for the locations
and properties of critical points get ever more complicated
the more the $SO(8)$ gauge symmetry is broken. As explained
in~\cite{Fischbacher:2010ki}, the large number of solutions suggests
using semi-automatic heuristics to obtain analytic expressions from
numerical data. We will now illustrate this approach   by 
proposing some analytic results  for critical
point \#9 in the list presented in~\cite{Fischbacher:2009cj}.  This solution 
has residual $U(1)\times U(1)$ gauge symmetry but as we will see, the analytic form of
the solution is very complicated.     We then discuss  critical
point \#8, which has  {\em no} residual continuous symmetry, and argue that it is,
at present, completely out of reach of analytic methods.

\subsection{Critical point \#9}

Using the sensitivity back-propagation method with enhanced numerical
precision allows the determination of the cosmological constant of
critical point~\#9 to high accuracy with reasonable
computational effort. To 100~digits, this is:
\begin{equation}
  \begin{array}{lcl}
    \cP^{\#9}/g^2&\approx&-10.6747541829948937677979769359580616537222601245605\\
    &&\phantom{-10.}8812167158899484460464464439307664949292077482553.
  \end{array}
\end{equation}

Employing the PSLQ algorithm~\cite{PSLQ} to find integer relations
between integer powers of $\cP^{\#9}/g^2$, one finds that about 80
digits of the cosmological constant suffice to automatically derive
the conjecture that $\left(\cP^{\#9}/g^2\right)^4$
is a zero of the polynomial
\begin{equation}
  x\mapsto -27x^3+351\,632x^2-13\,574\,400x+405\,504.
\end{equation}

Specifically, if we define:
\begin{equation}
\begin{array}{lcl}
Q&=&6561\\
R&=&28482192\\
S&=&122545537024\\
W&=&((128692865796145152+20596696547328\sqrt{2343}\,i)/1594323)^{1/3}\,, 
\end{array}
\end{equation}
then we find:
\begin{equation}
V^{\#9}/g^2 ~=~ \left(\frac{QW^2+RW+S}{QW}\right)^{1/4}
\end{equation}
It is also interesting to note that there is no natural number $N<100000$ for which 
$\left(W/|W|\right)^N=1$.

This exact expression then manages to `predict' the next 20 digits
correctly and hence very likely is the correct one. The complexity of
the analytic expression shown here has to be seen in contrast to those
that give the cosmological constant for the long-known critical points
\#1 -- \#7: There, the fourth power of $-\cP/g^2$ always is a fairly
manageable rational number.

Given that $\cP^{\#9}/g^2$ is a zero of a $12^{\rm th}$ order
polynomial with coefficients of magnitude $<10^9$, one may
optimistically hope that a reasonable number of digits of accuracy
should suffice to automatically determine similar analytic conjectures
for the entries of the $E_{7(7)}$ $56-$bein~$\cV$. However, as table
(A.1) in~\cite{Bobev:2010ib} shows, the coordinates of stationary
points generally can be expected to be algebraically somewhat more
complicated than their cosmological constants, and this seems to hold
as well for the entries of the $56$-bein. So far, 150 digits of
accuracy turned out to be insufficient to obtain any analytic
conjectures for this critical point by applying the PSLQ
algorithm. Still, once analytic expressions for the entries 
of~$\cV$ are known, exact analytic results for all other 
properties can be derived automatically. In
particular, the $Q^{ijkl}=0$ stationarity condition then can be checked
analytically. The same claims also hold for other stationary points
and for models in other dimensions, however, taking the logarithm of~$\cV$
and also establishing that~$\cV$ is indeed  an element of the
corresponding exceptional group may sometimes be somewhat tricky
(especially in three dimensions). For critical point \#9, the 
$56$-bein~$\cV$ contains (in its real and imaginary part) 92 different
irrational numbers for which analytic expressions yet have to be
found. Nevertheless, the observation that the stationarity condition
could be satisfied to more than~$100$ digits leaves little room for
doubt about the existence of this particular solution.

Precision data on the location of solution \#9 are listed in Appendix~\ref{numlocations}.

\subsection{Critical point \#8}

The hybrid analytic/numerical heuristics based on the PSLQ algorithm
can be applied to all the other critical points but, at present, this
algorithm does not appear to be strong enough to obtain any analytic
conjecture for properties of the least symmetric points. For example,
at critical point \#8 in \cite{Fischbacher:2009cj} the gauge group is
broken completely and this point illustrates the magnitude of the
analytic challenge.

Among those critical points without residual symmetry,
one would naturally expect the one with smallest $-\cP/g^2$ to have
the simplest analytic expressions. The value of the cosmological
constant (accurate to 300 digits) is:
\begin{equation}
  \begin{array}{lcl}
    \cP^{\#8}/g^2&\approx&-10.434712595009226792428131507556048070465670084352\\
            &&\phantom{-10.}032231183431856229077695992678217278594478357784\\
            &&\phantom{-10.}662335885763784608491863855940772595618694726435\\
            &&\phantom{-10.}652924643716227013639950308762502906095914068261\\
            &&\phantom{-10.}215830900373768835938911510578959286864096501271\\
            &&\phantom{-10.}029884695034616715155818453427676270759147128290\\
            &&\phantom{-10.}476481455688
  \end{array}
\end{equation}

This information should be sufficient to obtain a useful analytic
conjecture for~$\cP/g^2$ if it is the zero of a polynomial with
integer coefficients such that
\[\langle\mbox{degree}\rangle\cdot\langle\mbox{number of digits in max coeff}\rangle<300-n,\quad n\approx20.
\]

So far, we have not managed to obtain such an expression,
indicating that the complexity of the corresponding polynomial is
beyond reach even of this level of accuracy. We give precision data
on the location of critical point~$\#8$ to 150 digits in
Appendix~\ref{numlocations} and leave it as a challenge to our readers
to obtain an analytic result. 

This investigation shows that, in all likelihood, analytic expressions
for further critical points (which are very likely to exist) with no
or very little residual symmetry and large~$\cP/\cP_0$ probably are
too complicated to be handled conveniently. Hence, using numerical
methods may well be the dominant strategy to obtain information about
the properties of most as yet unknown critical points.

\section{Conclusions}
\label{Concs}

It is evident that, even many years after its construction, the potential of 
gauged $\Neql 8$ supergravity in four dimensions still has the capacity to
surprise us with new challenges and potentially physically important 
solutions.  

In this paper we have not only analytically exhibited a new, $\Neql 1$ supersymmetric critical point but we have embedded it in a consistent truncation to an $\Neql 1$ supersymmetric field theory.  We have given the complete superpotential for this theory and have derived the supersymmetric flow equations and used them to construct the flow from the $\Neql 8$ supersymmetric critical point  to the new $\Neql 1$ point.  The holographic interpretation of this flow is potentially very interesting because it appears to correspond to a new superconformal state in a broad class of flows whose original infra-red limit corresponds to a free fermion model \cite{Lin:2004nb}.

We have also done some analysis of the new families of solutions that have been found numerically  \cite{Fischbacher:2009cj, Fischbacher:2010ki}.  In addition to the discovery of the new $\Neql 1$ supersymmetric point,  the numerical search also revealed thirteen new non-supersymmetric critical points.  Unfortunately,  all of these new points have scalar excitations that violate the BF bound and so are unstable.  On the other hand, our complete stability analysis revealed that the $SO(3) \times SO(3)$-invariant critical point, found long ago, is perturbatively stable.  It is also a critical point that lies  in the truncation to the $\Neql 1$ supergravity theory and so it is easily analyzed from within a simple and very interesting supergravity theory.  It is also intriguing to note that this $\Neql 1$ supergravity has the somewhat unusual property that its potential has several critical points and {\it all}  of them are stable! This could possibly be related to its interesting holographic dual.

One of the other things that is clear from the numerical analysis is that there are probably a lot more critical points in the supergravity potential.  Moreover, a lot of them have little, or no, residual symmetry. We have demonstrated how such critical points with low levels of symmetry are going to be very hard, if not impossible, to access analytically.   On the other hand, the best strategy is probably the one exemplified by this paper:  Use the numerical algorithms to find the interesting, stable points and then, once one knows where to look, bring the full force of analytic methods to bear on the new solutions.

As regards the future, there is evidently a lot of new critical points to be discovered and one should also look in the five-dimensional maximal supergravity theories as well.  From the point of view of $AdS/CMT$, there are also a number of interesting new things to be done.

First, as we pointed out in Section \ref{contrunc},  the $\Neql 1$ supergravity that contains the  new, $\Neql 1$ supersymmetric critical point can itself be embedded in slightly larger $\Neql 2$ supergravity theories that contain some vector potentials.   These vector potentials can be used to induce chemical potentials in the theory on the brane and so drive interesting flows to the critical points of the theory, particularly given the stability of all the critical points in the  $\Neql 1$ theory.  This would be especially interesting given the well-known holographic interpretation of the massive flows within this model.

It is also important to point out that the massive flows with sixteen supersymmetries in gauged supergravity studied in \cite{Pope:2003jp} represented an extremely simple ``tip of an iceberg'' when the corresponding flows were studied in much greater generality in M-theory and IIB supergravity \cite{Bena:2004jw,Lin:2004nb}.  Since the flows studied here differ from the flows of \cite{Pope:2003jp} by adding perturbations by normalizable modes, it is reasonable to expect that something similar might occur in the far more general flows constructed in  \cite{Bena:2004jw,Lin:2004nb}.  Thus it would be extremely interesting to lift the solutions considered here to solutions of M-theory.  On the face of it, this is a dauntingly difficult problem because of the lack of symmetry:  Only a $U(1) \times U(1)$.  On the other hand, there might be other aspects to the geometry that could make the computations feasible, particularly if the underlying manifolds were Sasaki-Einstein, or something similar.  As yet, we have no basis for belief, one way or the other, in such a geometric simplification, but we think that this new supersymmetric phase is sufficiently interesting that it warrants a great deal of effort in trying to characterize its underlying geometry in as simple and universal manner as possible.

\bigskip
\bigskip
\leftline{\bf Acknowledgements}
\smallskip
The work of KP and NW  was supported in part by DOE grant DE-FG03-84ER-40168.  


\vfill\eject
\appendix
\renewcommand{\thesection}{\Alph{section}}
\renewcommand{\theequation}{A.\arabic{equation}}
\renewcommand\thetable{A.\arabic{table}}
\setcounter{equation}{0}
\setcounter{table}{0}
\label{appendixA}


\section{The scalar potential in the $ \sotsotgp $ sector}
\label{newpotso3}
The scalar potential in the $\sotsotgp$ sector is a function of   an $O(1,1)$ group element, 
$\rho = e^{-2\lambda_0/\sqrt 3}\,,
$
and a symmetric  $SL(3,\RR)$ matrix, $M=(m_{ij})$. We fix the gauge by setting  $m_{23}=0$ and   explicitly solve the constraint, $\det M=1$, 
\begin{equation}\label{}
m_{33}\eql {1\over\Delta}(1+m_{13}^2m_{22}^2)\,,\qquad \Delta\eql m_{11}m_{22}-m_{12}^2\,,
\end{equation}
thereby eliminating $m_{33}$. The resulting potential depends on five parameters: $\rho$, $m_{11}$, $m_{12}$, $m_{22}$ and $m_{13}$, and is given by
\begin{equation}\label{so3pot}
\cP\eql {g^2\over 8\Delta^2}\,\left[\,3\rho^{-1}P_{-1}+6\rho\,P_{1}+\rho^3 P_3\,\right]\,,
\end{equation}
where
\begin{equation}\label{}
\begin{split}
P_{-1} =  & -1-6 m_{ 12 }^4-m_{ 12 }^8-m_{ 12 }^4 m_{ 13 }^2-m_{ 12 }^6
   m_{ 13 }^2+12 m_{ 11 } m_{ 12 }^2 m_{ 22 } +4 m_{ 11 } m_{ 12 }^6
   m_{ 22 } \\ &  -2 m_{ 13 }^2 m_{ 22 }+2 m_{ 11 } m_{ 12 }^2 m_{ 13 }^2
   m_{ 22 }+2 m_{ 11 } m_{ 12 }^4 m_{ 13 }^2 m_{ 22 }-6 m_{ 11 }^2
   m_{ 22 }^2-6 m_{ 11 }^2 m_{ 12 }^4 m_{ 22 }^2  \\ & -m_{ 11 }^2
   m_{ 13 }^2 m_{ 22 }^2-m_{ 11 }^2 m_{ 12 }^2 m_{ 13 }^2
   m_{ 22 }^2-m_{ 12 }^4 m_{ 13 }^2 m_{ 22 }^2-m_{ 13 }^4
   m_{ 22 }^2+4 m_{ 11 }^3 m_{ 12 }^2 m_{ 22 }^3  \\ & +2 m_{ 11 }
   m_{ 12 }^2 m_{ 13 }^2 m_{ 22 }^3-m_{ 11 }^4
   m_{ 22 }^4-m_{ 11 }^2 m_{ 13 }^2 m_{ 22 }^4\,,
\end{split}
\end{equation}

\begin{equation}\label{}
\begin{split}
P_1 = & -m_{ 1 1 }^2-2 m_{ 1 2 }^2-m_{ 1 1 }^2 m_{ 1 2 }^4-2
   m_{ 1 2 }^6-2 m_{ 1 1 } m_{ 1 2 }^2 m_{ 1 3 }^2-m_{ 1 2 }^4
   m_{ 1 3 }^2-m_{ 1 2 }^6 m_{ 1 3 }^2  \\ & -m_{ 1 2 }^4 m_{ 1 3 }^4   +2
   m_{ 11 }^3 m_{ 12 }^2 m_{ 22 }+4 m_{ 11 } m_{ 12 }^4 m_{ 22 }-4
   m_{ 12 }^2 m_{ 13 }^2 m_{ 22 }+2 m_{ 11 } m_{ 12 }^2 m_{ 13 }^2
   m_{ 22 } \\ & +2 m_{ 11 } m_{ 12 }^4 m_{ 13 }^2
   m_{ 22 }-m_{ 22 }^2-m_{ 11 }^4 m_{ 22 }^2  -2 m_{ 11 }^2
   m_{ 12 }^2 m_{ 22 }^2-m_{ 12 }^4 m_{ 22 }^2 -m_{ 11 }^2
   m_{ 13 }^2 m_{ 22 }^2 \\ &  -m_{ 11 }^2 m_{ 12 }^2 m_{ 13 }^2
   m_{ 22 }^2-m_{ 12 }^4 m_{ 13 }^2 m_{ 22 }^2-2 m_{ 12 }^2
   m_{ 13 }^4 m_{ 22 }^2+2 m_{ 11 } m_{ 12 }^2 m_{ 22 }^3 -2
   m_{ 13 }^2 m_{ 22 }^3 \\ &  +2 m_{ 11 } m_{ 12 }^2 m_{ 13 }^2
   m_{ 22 }^3-m_{ 11 }^2 m_{ 22 }^4-m_{ 11 }^2 m_{ 13 }^2
   m_{ 22 }^4-m_{ 13 }^4 m_{ 22 }^4\,,
\end{split}
\end{equation}

\begin{equation}\label{}
\begin{split}
 P_3= &\  m_{ 11 }^4+4 m_{ 11 }^2 m_{ 12 }^2+m_{ 11 }^2
   m_{ 13 }^2+m_{ 12 }^2 m_{ 13 }^2+2 m_{ 11 }^3 m_{ 12 }^2
   m_{ 13 }^2+4 m_{ 11 } m_{ 12 }^4 m_{ 13 }^2 \\ & +m_{ 11 }^2
   m_{ 12 }^6 m_{ 13 }^2+m_{ 12 }^8 m_{ 13 }^2+2 m_{ 11 }
   m_{ 12 }^2 m_{ 13 }^4+m_{ 11 }^2 m_{ 12 }^4 m_{ 13 }^4+2
   m_{ 12 }^6 m_{ 13 }^4 \\ & +m_{ 12 }^4 m_{ 13 }^6+8 m_{ 11 }
   m_{ 12 }^2 m_{ 22 }+4 m_{ 11 }^2 m_{ 12 }^2 m_{ 13 }^2
   m_{ 22 }+6 m_{ 12 }^4 m_{ 13 }^2 m_{ 22 } \\ & -2 m_{ 11 }^3
   m_{ 12 }^4 m_{ 13 }^2 m_{ 22 }+2 m_{ 12 }^2 m_{ 13 }^4
   m_{ 22 }-2 m_{ 11 }^2 m_{ 22 }^2+4 m_{ 12 }^2 m_{ 22 }^2 \\ & +6
   m_{ 11 } m_{ 12 }^2 m_{ 13 }^2 m_{ 22 }^2+m_{ 11 }^4 m_{ 12 }^2
   m_{ 13 }^2 m_{ 22 }^2-3 m_{ 11 }^2 m_{ 12 }^4 m_{ 13 }^2
   m_{ 22 }^2  +3 m_{ 12 }^6 m_{ 13 }^2 m_{ 22 }^2 \\ & +2 m_{ 11 }^2
   m_{ 12 }^2 m_{ 13 }^4 m_{ 22 }^2+4 m_{ 12 }^4 m_{ 13 }^4
   m_{ 22 }^2+m_{ 12 }^2 m_{ 13 }^6 m_{ 22 }^2+8 m_{ 12 }^2
   m_{ 13 }^2 m_{ 22 }^3 \\ & +2 m_{ 11 }^3 m_{ 12 }^2 m_{ 13 }^2
   m_{ 22 }^3-6 m_{ 11 } m_{ 12 }^4 m_{ 13 }^2 m_{ 22 }^3+2
   m_{ 11 } m_{ 12 }^2 m_{ 13 }^4 m_{ 22 }^3+m_{ 22 }^4 \\ & +3
   m_{ 11 }^2 m_{ 12 }^2 m_{ 13 }^2 m_{ 22 }^4+m_{ 12 }^4
   m_{ 13 }^2 m_{ 22 }^4+4 m_{ 12 }^2 m_{ 13 }^4 m_{ 22 }^4+2
   m_{ 13 }^2 m_{ 22 }^5\\ & -2 m_{ 11 } m_{ 12 }^2 m_{ 13 }^2
   m_{ 22 }^5+m_{ 11 }^2 m_{ 13 }^2 m_{ 22 }^6+m_{ 13 }^4
   m_{ 22 }^6\,.
\end{split}
\end{equation}

\vfill\eject
\renewcommand{\theequation}{C.\arabic{equation}}
\renewcommand\thetable{B.\arabic{table}}
\setcounter{equation}{0}
\setcounter{table}{0}
\label{appendixC}

\section{Numerical data}

\subsection{Mass matrices}
\label{nummasses}

For convenience, numerical data on scalar masses are listed in
table~\ref{tbl:scalarmasses}. These values have been obtained
by \eqref{scalmassmat} and verified independently via taking second
order derivatives numerically. The labeling of stationary points
parallels \cite{Fischbacher:2009cj}, while the $AdS$ mass scale
conventions match those used in~\cite{Bobev:2010ib}. As the actual
position of most new critical points is only known to limited
accuracy, it is conceivable that better location data would in some
cases give very slightly different masses. Still, these numerical data
clearly show that only one of the 17 non-supersymmetric critical
points that are now known is perturbatively stable.


\def\massblockZ{\hline
$m^2_{\rm scalar}L^2$\quad\begin{minipage}{10.5cm}
{\vskip0,25ex
\small\tt
{\tt -2.000}$_{\times 70}$
}\end{minipage}\\
\hline
}

\def\massblockZN{\hline
$m^2_{\rm scalar}L^2$\quad\begin{minipage}{10.5cm}
{\vskip0,25ex
\small\tt
{\tt -2.400}$_{\times 27}$, {\tt -1.200}$_{\times 35}$, {\tt 0.000}$_{\times 7}$, {\tt 6.000}
}\end{minipage}\\
\hline
}

\def\massblockZNN{\hline
$m^2_{\rm scalar}L^2$\quad\begin{minipage}{10.5cm}
{\vskip0,25ex
\small\tt
{\tt -2.400}$_{\times 27}$, {\tt -1.200}$_{\times 35}$, {\tt 0.000}$_{\times 7}$, {\tt 6.000}
}\end{minipage}\\
\hline
}

\def\massblockZNNN{\hline
$m^2_{\rm scalar}L^2$\quad\begin{minipage}{10.5cm}
{\vskip0,25ex
\small\tt
{\tt -2.242}$_{\times 27}$, {\tt -1.425}$_{\times 27}$, {\tt 0.000}$_{\times 14}$, {\tt 1.550}, {\tt 6.449}
}\end{minipage}\\
\hline
}

\def\massblockZNNNN{\hline
$m^2_{\rm scalar}L^2$\quad\begin{minipage}{10.5cm}
{\vskip0,25ex
\small\tt
{\tt -2.222}$_{\times 12}$, {\tt -2.000}$_{\times 16}$, {\tt -1.556}$_{\times 18}$, {\tt -1.123}, {\tt 0.000}$_{\times 19}$, {\tt 2.000}$_{\times 3}$, {\tt 7.123}
}\end{minipage}\\
\hline
}

\def\massblockZNNNNN{\hline
$m^2_{\rm scalar}L^2$\quad\begin{minipage}{10.5cm}
{\vskip0,25ex
\small\tt
{\tt -3.000}$_{\times 20}$, {\tt -0.750}$_{\times 20}$, {\tt 0.000}$_{\times 28}$, {\tt 6.000}$_{\times 2}$
}\end{minipage}\\
\hline
}

\def\massblockZNNNNNN{\hline
$m^2_{\rm scalar}L^2$\quad\begin{minipage}{10.5cm}
{\vskip0,25ex
\small\tt
{\tt -1.714}$_{\times 20}$,
{\tt -1.312}$_{\times 9}$,
{\tt 0.000}$_{\times 22}$,
{\tt 2.571}$_{\times 9}$,
{\tt 5.598}$_{\times 9}$,
{\tt 8.571}$_{\times 1}$
}\end{minipage}\\
\hline
}

\def\massblockZNNNNNNN{\hline
$m^2_{\rm scalar}L^2$\quad\begin{minipage}{10.5cm}
{\vskip0,25ex
\small\tt
{\tt -3.051}, {\tt -2.476}, {\tt -2.433}$_{\times 2}$, {\tt -2.197}$_{\times 2}$, {\tt -2.094}, {\tt -1.968}$_{\times 2}$, {\tt -1.815}$_{\times 2}$, {\tt -1.814}$_{\times 2}$, {\tt -1.791}$_{\times 2}$, {\tt -1.768}$_{\times 2}$, {\tt -1.764}$_{\times 2}$, {\tt -1.606}, {\tt -1.398}$_{\times 2}$, {\tt -1.349}, {\tt -1.332}$_{\times 2}$, {\tt -1.330}$_{\times 2}$, {\tt -1.279}$_{\times 2}$, {\tt -1.268}$_{\times 2}$, {\tt -1.007}, {\tt -0.662}$_{\times 2}$, {\tt 0.000}$_{\times 27}$, {\tt 1.732}, {\tt 4.520}, {\tt 4.884}$_{\times 2}$, {\tt 5.576}$_{\times 2}$, {\tt 7.355}, {\tt 7.486}$_{\times 2}$
}\end{minipage}\\
\hline
}

\def\massblockZNNNNNNNN{\hline
$m^2_{\rm scalar}L^2$\quad\begin{minipage}{10.5cm}
{\vskip0,25ex
\small\tt
{\tt -3.076}, {\tt -2.598}, {\tt -2.568}, {\tt -2.451}$_{\times 2}$, {\tt -2.407}, {\tt -2.349}, {\tt -2.118}, {\tt -1.874}, {\tt -1.863}, {\tt -1.847}$_{\times 2}$, {\tt -1.826}, {\tt -1.792}$_{\times 2}$, {\tt -1.736}$_{\times 2}$, {\tt -1.699}, {\tt -1.691}, {\tt -1.526}$_{\times 2}$, {\tt -1.377}, {\tt -1.352}, {\tt -1.189}$_{\times 2}$, {\tt -1.166}, {\tt -1.158}, {\tt -1.027}, {\tt -0.971}, {\tt -0.640}, {\tt 0.000}$_{\times 28}$, {\tt 0.496}, {\tt 1.295}, {\tt 3.516}, {\tt 4.074}, {\tt 4.205}$_{\times 2}$, {\tt 5.378}, {\tt 6.156}$_{\times 2}$, {\tt 7.380}, {\tt 7.462}, {\tt 7.937}
}\end{minipage}\\
\hline
}

\def\massblockZNNNNNNNNN{\hline
$m^2_{\rm scalar}L^2$\quad\begin{minipage}{10.5cm}
{\vskip0,25ex
\small\tt
{\tt -3.367}, {\tt -3.292}$_{\times 2}$, {\tt -2.638}, {\tt -2.472}, {\tt -2.132}$_{\times 2}$, {\tt -2.086}$_{\times 4}$, {\tt -1.924}$_{\times 2}$, {\tt -1.696}$_{\times 2}$, {\tt -1.635}$_{\times 4}$, {\tt -1.282}$_{\times 2}$, {\tt -1.241}$_{\times 2}$, {\tt -1.021}$_{\times 2}$, {\tt -0.780}, {\tt -0.748}, {\tt -0.040}$_{\times 2}$, {\tt 0.000}$_{\times 26}$, {\tt 0.987}$_{\times 4}$, {\tt 2.132}, {\tt 3.814}$_{\times 2}$, {\tt 4.329}$_{\times 2}$, {\tt 5.762}$_{\times 2}$, {\tt 7.146}, {\tt 7.898}, {\tt 7.899}, {\tt 9.689}
}\end{minipage}\\
\hline
}

\def\massblockZNNNNNNNNNN{\hline
$m^2_{\rm scalar}L^2$\quad\begin{minipage}{10.5cm}
{\vskip0,25ex
\small\tt
{\tt -3.515}$_{\times 5}$, {\tt -2.485}, {\tt -2.121}$_{\times 4}$, {\tt -1.409}$_{\times 4}$, {\tt -1.286}, {\tt -1.092}$_{\times 2}$, {\tt -0.783}$_{\times 4}$, {\tt -0.515}$_{\times 8}$, {\tt 0.000}$_{\times 26}$, {\tt 2.485}, {\tt 3.268}$_{\times 4}$, {\tt 4.801}, {\tt 5.652}$_{\times 4}$, {\tt 6.364}$_{\times 2}$, {\tt 7.029}, {\tt 8.485}$_{\times 2}$
}\end{minipage}\\
\hline
}

\def\massblockZNNNNNNNNNNN{\hline
$m^2_{\rm scalar}L^2$\quad\begin{minipage}{10.5cm}
{\vskip0,25ex
\small\tt
{\tt -2.250}$_{\times 4}$, {\tt -2.229}$_{\times 4}$, {\tt -2.196}, {\tt -2.000}$_{\times 4}$, {\tt -1.912}$_{\times 4}$, {\tt -1.517}$_{\times 4}$, {\tt -1.250}$_{\times 4}$, {\tt -0.732}, {\tt 0.000}$_{\times 30}$, {\tt 2.732}, {\tt 3.067}$_{\times 3}$, {\tt 3.067}, {\tt 4.412}$_{\times 4}$, {\tt 8.196}, {\tt 8.679}$_{\times 4}$
}\end{minipage}\\
\hline
}

\def\massblockZNNNNNNNNNNNN{\hline
$m^2_{\rm scalar}L^2$\quad\begin{minipage}{10.5cm}
{\vskip0,25ex
\small\tt
{\tt -2.812}$_{\times 2}$, {\tt -2.714}$_{\times 2}$, {\tt -2.595}, {\tt -2.174}, {\tt -2.056}, {\tt -1.754}$_{\times 2}$, {\tt -1.488}$_{\times 2}$, {\tt -1.410}, {\tt -1.374}, {\tt -1.339}$_{\times 2}$, {\tt -1.232}$_{\times 2}$, {\tt -1.194}, {\tt -1.159}, {\tt -1.118}$_{\times 2}$, {\tt -0.527}, {\tt -0.480}$_{\times 2}$, {\tt -0.300}$_{\times 2}$, {\tt 0.000}$_{\times 27}$, {\tt 1.322}, {\tt 1.789}, {\tt 3.995}$_{\times 2}$, {\tt 4.024}, {\tt 4.668}$_{\times 2}$, {\tt 5.203}, {\tt 5.736}$_{\times 2}$, {\tt 5.861}, {\tt 7.007}$_{\times 2}$, {\tt 7.073}, {\tt 8.228}, {\tt 11.974}, {\tt 12.284}
}\end{minipage}\\
\hline
}
\def\massblockZNNNNNNNNNNNNN{\hline
$m^2_{\rm scalar}L^2$\quad\begin{minipage}{10.5cm}
{\vskip0,25ex
\small\tt
{\tt -3.007}, {\tt -2.998}, {\tt -2.621}$_{\times 2}$, {\tt -2.301}, {\tt -2.178}, {\tt -2.068}, {\tt -1.864}, {\tt -1.862}, {\tt -1.543}$_{\times 2}$, {\tt -1.437}, {\tt -1.389}, {\tt -1.319}$_{\times 2}$, {\tt -1.252}$_{\times 2}$, {\tt -1.216}, {\tt -1.101}, {\tt -1.071}, {\tt -0.973}, {\tt -0.903}, {\tt -0.102}, {\tt 0.000}$_{\times 28}$, {\tt 0.391}, {\tt 1.073}$_{\times 2}$, {\tt 1.257}, {\tt 2.339}, {\tt 3.688}, {\tt 4.914}$_{\times 2}$, {\tt 5.471}, {\tt 5.549}$_{\times 2}$, {\tt 5.984}, {\tt 6.109}, {\tt 6.991}, {\tt 8.135}, {\tt 8.531}, {\tt 10.188}$_{\times 2}$, {\tt 11.360}
}\end{minipage}\\
\hline
}

\def\massblockZNNNNNNNNNNNNNN{\hline
$m^2_{\rm scalar}L^2$\quad\begin{minipage}{10.5cm}
{\vskip0,25ex
\small\tt
{\tt -2.857}, {\tt -2.754}$_{\times 2}$, {\tt -2.430}$_{\times 2}$, {\tt -2.130}, {\tt -1.971}, {\tt -1.842}$_{\times 2}$, {\tt -1.795}, {\tt -1.451}$_{\times 2}$, {\tt -1.006}, {\tt -0.939}$_{\times 2}$, {\tt -0.884}$_{\times 2}$, {\tt -0.849}, {\tt -0.741}$_{\times 2}$, {\tt 0.000}$_{\times 27}$, {\tt 0.257}$_{\times 2}$, {\tt 1.063}, {\tt 1.884}, {\tt 2.892}$_{\times 2}$, {\tt 3.311}$_{\times 2}$, {\tt 3.937}$_{\times 2}$, {\tt 4.303}, {\tt 5.533}, {\tt 7.717}, {\tt 7.746}$_{\times 2}$, {\tt 8.018}$_{\times 2}$, {\tt 8.580}, {\tt 8.683}, {\tt 17.599}, {\tt 17.612}, {\tt 20.534}, {\tt 20.564}
}\end{minipage}\\
\hline
}

\def\massblockZNNNNNNNNNNNNNNN{\hline
$m^2_{\rm scalar}L^2$\quad\begin{minipage}{10.5cm}
{\vskip0,25ex
\small\tt
{\tt -2.975}, {\tt -2.730}, {\tt -2.667}, {\tt -1.807}, {\tt -1.790}, {\tt -1.772}, {\tt -1.593}, {\tt -1.568}, {\tt -1.550}, {\tt -1.548}, {\tt -1.345}, {\tt -1.267}, {\tt -1.245}, {\tt -1.119}, {\tt -0.907}, {\tt -0.846}, {\tt -0.793}, {\tt -0.430}, {\tt 0.000}$_{\times 28}$, {\tt 0.338}, {\tt 0.653}, {\tt 1.439}, {\tt 1.991}, {\tt 2.335}, {\tt 2.446}, {\tt 2.470}, {\tt 2.916}, {\tt 4.174}, {\tt 4.454}, {\tt 5.969}, {\tt 6.310}, {\tt 7.583}, {\tt 7.920}, {\tt 7.945}, {\tt 7.985}, {\tt 8.175}, {\tt 8.905}, {\tt 9.267}, {\tt 9.762}, {\tt 10.891}, {\tt 11.208}, {\tt 14.959}, {\tt 15.076}
}\end{minipage}\\
\hline
}

\def\massblockZNNNNNNNNNNNNNNNN{\hline
$m^2_{\rm scalar}L^2$\quad\begin{minipage}{10.5cm}
{\vskip0,25ex
\small\tt
{\tt -3.470}, {\tt -3.196}, {\tt -3.021}, {\tt -2.739}, {\tt -2.335}, {\tt -1.826}, {\tt -1.695}, {\tt -1.641}, {\tt -1.403}, {\tt -1.252}, {\tt -1.088}, {\tt -1.076}, {\tt -0.714}, {\tt -0.665}, {\tt -0.475}, {\tt 0.000}$_{\times 28}$, {\tt 0.043}, {\tt 0.555}, {\tt 1.418}, {\tt 2.490}, {\tt 2.508}, {\tt 3.069}, {\tt 3.231}, {\tt 3.549}, {\tt 3.865}, {\tt 4.493}, {\tt 4.913}, {\tt 4.951}, {\tt 5.148}, {\tt 5.710}, {\tt 6.431}, {\tt 7.783}, {\tt 7.855}, {\tt 7.929}, {\tt 7.972}, {\tt 8.082}, {\tt 8.863}, {\tt 9.845}, {\tt 11.638}, {\tt 11.712}, {\tt 12.309}, {\tt 18.666}, {\tt 18.751}
}\end{minipage}\\
\hline
}

\def\massblockZNNNNNNNNNNNNNNNNN{\hline
$m^2_{\rm scalar}L^2$\quad\begin{minipage}{10.5cm}
{\vskip0,25ex
\small\tt
{\tt -3.358}, {\tt -3.291}, {\tt -2.830}, {\tt -2.776}, {\tt -2.307}, {\tt -2.274}, {\tt -1.994}$_{\times 2}$, {\tt -1.898}, {\tt -1.150}$_{\times 2}$, {\tt -0.945}, {\tt -0.796}, {\tt -0.724}, {\tt -0.686}, {\tt -0.626}, {\tt -0.335}$_{\times 2}$, {\tt 0.000}$_{\times 28}$, {\tt 1.647}, {\tt 2.012}$_{\times 2}$, {\tt 2.290}, {\tt 4.228}, {\tt 4.270}$_{\times 2}$, {\tt 5.563}, {\tt 6.342}$_{\times 2}$, {\tt 6.656}, {\tt 6.896}, {\tt 7.209}, {\tt 7.323}, {\tt 7.358}$_{\times 2}$, {\tt 7.652}, {\tt 8.033}$_{\times 2}$, {\tt 8.269}, {\tt 10.677}, {\tt 10.966}, {\tt 15.802}, {\tt 15.892}
}\end{minipage}\\
\hline
}

\def\massblockZNNNNNNNNNNNNNNNNNN{\hline
$m^2_{\rm scalar}L^2$\quad\begin{minipage}{10.5cm}
{\vskip0,25ex
\small\tt
{\tt -4.227}, {\tt -3.671}, {\tt -3.029}, {\tt -2.320}$_{\times 2}$, {\tt -1.955}, {\tt -1.804}, {\tt -1.267}, {\tt -1.252}, {\tt -0.746}$_{\times 2}$, {\tt -0.391}, {\tt -0.364}, {\tt 0.000}$_{\times 28}$, {\tt 1.286}, {\tt 3.114}, {\tt 3.442}, {\tt 3.871}, {\tt 4.092}, {\tt 4.180}, {\tt 4.748}$_{\times 2}$, {\tt 4.820}$_{\times 2}$, {\tt 6.519}, {\tt 7.816}, {\tt 7.922}, {\tt 8.240}, {\tt 9.034}$_{\times 2}$, {\tt 9.127}, {\tt 10.227}$_{\times 2}$, {\tt 11.045}, {\tt 11.179}, {\tt 11.535}, {\tt 11.968}$_{\times 2}$, {\tt 14.471}, {\tt 18.868}, {\tt 20.926}$_{\times 2}$, {\tt 22.287}
}\end{minipage}\\
\hline
}

\def\massblockZNNNNNNNNNNNNNNNNNNN{\hline
$m^2_{\rm scalar}L^2$\quad\begin{minipage}{10.5cm}
{\vskip0,25ex
\small\tt
{\tt -4.447}, {\tt -3.918}, {\tt -3.905}, {\tt -2.896}, {\tt -2.408}, {\tt -1.007}$_{\times 2}$, {\tt -0.591}, {\tt -0.468}, {\tt -0.246}, {\tt -0.093}$_{\times 2}$, {\tt -0.004}, {\tt 0.000}$_{\times 28}$, {\tt 0.657}$_{\times 2}$, {\tt 2.255}, {\tt 3.408}, {\tt 3.743}, {\tt 4.224}, {\tt 4.282}, {\tt 4.765}, {\tt 4.782}, {\tt 5.812}$_{\times 2}$, {\tt 6.202}$_{\times 2}$, {\tt 6.438}$_{\times 2}$, {\tt 7.719}, {\tt 9.859}, {\tt 10.305}, {\tt 10.614}, {\tt 12.377}$_{\times 2}$, {\tt 12.537}$_{\times 2}$, {\tt 14.437}, {\tt 14.464}, {\tt 16.669}, {\tt 16.748}, {\tt 28.669}, {\tt 28.685}
}\end{minipage}\\
\hline
}

\def\massblockZNNNNNNNNNNNNNNNNNNNN{\hline
$m^2_{\rm scalar}L^2$\quad\begin{minipage}{10.5cm}
{\vskip0,25ex
\small\tt
{\tt -3.089}, {\tt -2.400}, {\tt -2.166}, {\tt -1.825}, {\tt -1.744}, {\tt -1.481}, {\tt -1.316}, {\tt -1.268}, {\tt -0.941}, {\tt -0.670}, {\tt -0.489}, {\tt 0.000}$_{\times 28}$, {\tt 0.750}, {\tt 1.379}, {\tt 1.605}, {\tt 1.759}, {\tt 1.920}, {\tt 3.345}, {\tt 3.382}, {\tt 3.421}, {\tt 3.822}, {\tt 4.931}, {\tt 5.930}, {\tt 6.725}, {\tt 7.256}, {\tt 7.969}, {\tt 7.998}, {\tt 8.049}, {\tt 8.210}, {\tt 9.836}, {\tt 10.366}, {\tt 10.824}, {\tt 11.837}, {\tt 12.629}, {\tt 12.725}, {\tt 13.048}, {\tt 13.863}, {\tt 18.891}, {\tt 19.022}, {\tt 24.203}, {\tt 24.239}, {\tt 25.656}, {\tt 25.669}
}\end{minipage}\\
\hline
}

\begin{longtable}{|l|}
\hline
\raise-0.1ex\hbox{\small {\bf \#0}: $\cP/g^2=-6.000000$, $\mathcal{N}=8$}\\
\massblockZ
\hline\hline
\raise-0.1ex\hbox{\small {\bf \#1}: $\cP/g^2=-6.687403$, $SO(7)_+$}\\
\massblockZN
\hline\hline
\raise-0.1ex\hbox{\small {\bf \#2}: $\cP/g^2=-6.987712$, $SO(7)_-$}\\
\massblockZNN
\hline\hline
\raise-0.1ex\hbox{\small {\bf \#3}: $\cP/g^2=-7.191576$, $G_2$ $\mathcal{N}=1$}\\
\massblockZNNN
\hline\hline
\raise-0.1ex\hbox{\small {\bf \#4}: $\cP/g^2=-7.794229$, $SU(3)\times U(1)$ $\mathcal{N}=2$}\\
\massblockZNNNN
\hline\hline
\raise-0.1ex\hbox{\small {\bf \#5}: $\cP/g^2=-8.000000$, $SU(4)_-$}\\
\massblockZNNNNN
\hline\hline
\raise-0.1ex\hbox{\small {\bf \#6}: $\cP/g^2=-14.000000$, $SO(3)\times SO(3)$}\\
\massblockZNNNNNN
\hline\hline
\raise-0.1ex\hbox{\small {\bf \#7}: $\cP/g^2=-9.987083$, $U(1)$}\\
\massblockZNNNNNNN
\hline\hline
\raise-0.1ex\hbox{\small {\bf \#8}: $\cP/g^2=-10.434713$, \nosymmetry}\\
\massblockZNNNNNNNN
\hline\hline
\raise-0.1ex\hbox{\small {\bf \#9}: $\cP/g^2=-10.674754$, $U(1)\times U(1)$}\\
\massblockZNNNNNNNNN
\hline\hline
\raise-0.1ex\hbox{\small {\bf \#10}: $\cP/g^2=-11.656854$, $U(1)\times U(1)$}\\
\massblockZNNNNNNNNNN
\hline\hline
\raise-0.1ex\hbox{\small {\bf \#11}: $\cP/g^2=-12.000000$, $U(1)\times U(1)$ $\mathcal{N}=1$}\\
\massblockZNNNNNNNNNNN
\hline\hline
\raise-0.1ex\hbox{\small {\bf \#12}: $\cP/g^2=-13.623653$, $U(1)$}\\
\massblockZNNNNNNNNNNNN
\hline\hline
\raise-0.1ex\hbox{\small {\bf \#13}: $\cP/g^2=-13.676114$, \nosymmetry}\\
\massblockZNNNNNNNNNNNNN
\hline\hline
\raise-0.1ex\hbox{\small {\bf \#14}: $\cP/g^2=-14.970385$, $U(1)$}\\
\massblockZNNNNNNNNNNNNNN
\hline\hline
\raise-0.1ex\hbox{\small {\bf \#15}: $\cP/g^2=-16.414456$, \nosymmetry}\\
\massblockZNNNNNNNNNNNNNNN
\hline\hline
\raise-0.1ex\hbox{\small {\bf \#16}: $\cP/g^2=-17.876443$, \nosymmetry}\\
\massblockZNNNNNNNNNNNNNNNN
\hline\hline
\raise-0.1ex\hbox{\small {\bf \#17}: $\cP/g^2=-18.052693$, \nosymmetry}\\
\massblockZNNNNNNNNNNNNNNNNN
\hline\hline
\raise-0.1ex\hbox{\small {\bf \#18}: $\cP/g^2=-21.265976$, \nosymmetry}\\
\massblockZNNNNNNNNNNNNNNNNNN
\hline\hline
\raise-0.1ex\hbox{\small {\bf \#19}: $\cP/g^2=-21.408498$, \nosymmetry}\\
\massblockZNNNNNNNNNNNNNNNNNNN
\hline\hline
\raise-0.1ex\hbox{\small {\bf \#20}: $\cP/g^2=-25.149369$, \nosymmetry}\\
\massblockZNNNNNNNNNNNNNNNNNNNN
\hline
\caption{Scalar masses of the stationary points \#0 - \#20 listed
  in~\cite{Fischbacher:2009cj}, using the conventions of~\cite{Bobev:2010ib}\label{tbl:scalarmasses}}
\end{longtable}

\subsection{The   $\sotsotgp$  point}
\label{extrapoints}

The following table contains the numerical data giving  the location (in the conventions of \cite{Fischbacher:2009cj, Fischbacher:2010ki}),   continous symmetries, and   mass spectra of spin-3/2 and spin-1/2 fields   that were  used to set up an analytic calculation of the new $\sotsotgp$-invariant critical point in Section \ref{newso3point}.
\bigskip

\begin{longtable}{||l||}
\hline
\raise-0.1ex\hbox{\small {\bf Extremum}: $V/g^2=-8.807339$, Quality: $|Q|=10^{-11.84}$, $|\nabla|=10^{-4.45}$ }\\
\hline
$\phi$\quad\begin{minipage}{10.5cm}
{\vskip0.25ex
\small\tt
$-0.3299_{[1247]+}$, $+0.3299_{[1257]+}$, $-0.3299_{[1347]+}$, $-0.3299_{[1357]+}$, $-0.4666_{[1678]+}$, $-0.4666_{[2345]+}$, $-0.3299_{[2468]+}$, $+0.3299_{[2568]+}$, $-0.3299_{[3468]+}$, $-0.3299_{[3568]+}$, $+0.2866_{[1236]-}$, $+0.2026_{[1247]-}$, $+0.2026_{[1257]-}$, $-0.2026_{[1347]-}$, $+0.2026_{[1357]-}$, $+0.2866_{[1456]-}$, $-0.2866_{[2378]-}$, $-0.2026_{[2468]-}$, $-0.2026_{[2568]-}$, $+0.2026_{[3468]-}$, $-0.2026_{[3568]-}$, $-0.2866_{[4578]-}$
}\end{minipage}\\
\hline
Symmetry\quad\begin{minipage}{8.5cm}
{\vskip0.25ex
\small\tt
[6-dimensional]
}\end{minipage}\\
\hline

$(m^2/m_0^2)[\psi]$\quad\begin{minipage}{8.5cm}
{\vskip0.25ex
\small
{\tt 2.049}$_{(\times 8)}$
}\end{minipage}\\
\hline
$(m^2/m_0^2)[\chi]$\quad\begin{minipage}{8.5cm}
{\vskip0.25ex
\small
{\tt 4.098}$_{(\times 8)}$, {\tt 3.774}$_{(\times 8)}$, {\tt 0.391}$_{(\times 32)}$, {\tt 0.007}$_{(\times 8)}$
}\end{minipage}\\
\hline
\end{longtable}

\subsection{The locations of critical point \#8 and \#9}
\label{numlocations}

To 150 accurate digits, the $70$-vector $\phi_{\tt n}$ that gives the
$56$-bein for critical point~\#8 according to
\begin{equation}
{\mathcal{V}}^{\tilde{\mathcal{A}}}{}_{\tilde{\mathcal{B}}}=\exp\Big(\sum_{\tt n}\phi_{\tt n} g^{({\tt n})}\Big)^{\tilde{\mathcal{A}}}{}_{\tilde{\mathcal{B}}}
\end{equation}
(using the conventions of~\cite{Fischbacher:2009cj}) is listed below
-- only nonzero entries are given, and index counting starts at~1.
Furthermore, the following relations between entries have been 
employed to further shorten the presentation, i.e. entry $\phi_{02}$
is not listed as this is related in a simple way to $\phi_{01}$, etc.:
\begin{equation}\label{relationsV8}\begin{split}
&\quad \phi_{01} = \coeff{1}{2}  \phi_{02}\,,\qquad \phi_{06}\eql 2 \phi_{07}\,,\qquad \phi_{08}\eql \phi_{09}\eql -\phi_{16}\eql \phi_{21}\,,\qquad \phi_{10}\eql-\phi_{15}\,,
\\[6 pt]
&\qquad   \phi_{31}\eql -\phi_{34}\,,\qquad \phi_{32}\eql -\phi_{33}\,,\qquad \phi_{36}\eql \coeff{1}{2}\phi_{37}\eql \phi_{38}\eql \phi_{45}\eql\phi_{50}\,,\\[6 pt]
& \phi_{40}\eql \phi_{42}\eql\phi_{67}\eql -\phi_{68}\eql \coeff{1}{2}\phi_{41}\,,\qquad \phi_{44}\eql-\phi_{51}\,,\qquad \phi_{65}\eql -\phi_{66}\eql \phi_{69}\eql\phi_{70}\,.
\end{split}
\end{equation}

{\small\tt
\begin{longtable}{|lll|}
\hline
&&Critical point \#8\\
\hline
$\phi_{01}$&=&+0.08149800991420636144284179340047368401301087858848\\
&&\phantom{+0.}42800995230168482979337459751986503947341102536696\\
&&\phantom{+0.}59441836820594742957249335857968784123454183582584\\
\hline
$\phi_{03}$&=&-0.03314238242442091820006312903856728674314315582873\\
&&\phantom{+0.}33607526567504541404243330729534141329646921199538\\
&&\phantom{+0.}45271667847520130139998373725460397478551252545118\\
\hline
$\phi_{04}$&=&-0.22928078467725455928580984487808194151230806883443\\
&&\phantom{+0.}52817043595346048767161580963041290553976047472470\\
&&\phantom{+0.}09427009336229746194495419166858363204010872255404\\
\hline
$\phi_{05}$&=&-0.10445671578332632889643239570691855166585035939984\\
&&\phantom{+0.}09510559494413203503301949310072312320346077670660\\
&&\phantom{+0.}27840429065891749231499259158427087655108299991042\\
\hline
$\phi_{06}$&=&+0.02036735311060190149294505346424483818060735003475\\
&&\phantom{+0.}33795924606519641760557682342896665913283892131149\\
&&\phantom{+0.}53746151204446247731496900850004187893794272273321\\
\hline
$\phi_{08}$&=&-0.03559865621956356146272131345605921475734803407844\\
&&\phantom{+0.}80808453768177720686695293865553919293919403128531\\
&&\phantom{+0.}60885826226938208398313742092471961391351069371766\\
\hline
$\phi_{10}$&=&+0.13881820608352000126429425461999416939108789579709\\
&&\phantom{+0.}31005256129004995171127854992746826585835114404814\\
&&\phantom{+0.}11798589154652179505873190649683374924456901646435\\
\hline
$\phi_{30}$&=&+0.02677921177333006507310803029303476490770449118409\\
&&\phantom{+0.}72584501818311530093660524824522672899077535202704\\
&&\phantom{+0.}97803921727780281853237039415725239476439754669138\\
\hline
$\phi_{31}$&=&-0.06942962363246406699044756399786065315982260999419\\
&&\phantom{+0.}98966289903099323484896576064624324183385175162121\\
&&\phantom{+0.}52624041595162697366413873359035893406104209038188\\
\hline
$\phi_{32}$&=&-0.06750387272461459056792498795164290446838069222598\\
&&\phantom{+0.}55102223202096333072069236412208655595135957933692\\
&&\phantom{+0.}29229827936280560414372305216716684747899854200512\\
\hline
$\phi_{35}$&=&-0.16563845903825819905400315828875607122734971117249\\
&&\phantom{+0.}70517081624510177063453676953771321265847885526948\\
&&\phantom{+0.}03052004918105676586064786133797026288648172745514\\
\hline
$\phi_{36}$&=&+0.10378459150725177347878495867068630126872774257821\\
&&\phantom{+0.}35237695229651616198582397594753191694042112778283\\
&&\phantom{+0.}96565840528022497448214465709422149654857239137791\\
\hline
$\phi_{40}$&=&+0.12073223619836159819915224622226067361300930438839\\
&&\phantom{+0.}96035236216960804405540107886213345202334369260048\\
&&\phantom{+0.}27377925737213842511361330833997851495203509305436\\
\hline
$\phi_{43}$&=&+0.12684630664797953637045948007446536096364288507262\\
&&\phantom{+0.}81804893417909807481600218573392242444093258304550\\
&&\phantom{+0.}19824947445386212857704199239094827036377909217641\\
\hline
$\phi_{44}$&=&+0.03017961664646812668987531418299021421338719589454\\
&&\phantom{+0.}93792965565568997142878587951334137071978438388328\\
&&\phantom{+0.}46228107448721261587957827445688491766425382855899\\
\hline
$\phi_{56}$&=&-0.18720553994091578975021010844044578939041727686172\\
&&\phantom{+0.}69390824549047801767357394476060516588050135081207\\
&&\phantom{+0.}12281162342828736033619854130471810569228674929438\\
\hline
$\phi_{65}$&=&-0.11016546490484786642658912198932195287609915468539\\
&&\phantom{+0.}56525300344607258210753372221149767989284706041286\\
&&\phantom{+0.}20213656425808941886997690051805968358012094919480\\
\hline
\end{longtable}
}
\bigskip

To 130 digits, the corresponding $70$-vector for critical point \#9
is given in the following table, where we have made use of these
relations to shorten the presentation:

\begin{equation}\label{relationsV9}
\begin{array}{l}
\phi_{11}\eql-\phi_{12}\eql-\phi_{17}\eql\phi_{18}\,,\qquad\phi_{21}\eql\phi_{24}\eql\phi_{25}\eql\phi_{28}\eql\phi_{29}\,,\\[6 pt]
\phi_{36}\eql\phi_{42}\,,\qquad \phi_{37}\eql\phi_{41}\,,\qquad\phi_{38}\eql\phi_{40}\,,\\[6 pt]
\phi_{43}\eql-\phi_{46}\eql-\phi_{47}\eql-\phi_{52}\eql-\phi_{53}\eql-\phi_{56}\eql\phi_{59}\eql\\[6 pt]
\quad\,\,\,\eql-\phi_{60}\eql-\phi_{63}\eql\phi_{64}\eql-\phi_{65}\eql\phi_{70}.
\end{array}
\end{equation}
\vfill\eject

{\small\tt
\begin{longtable}{|lll|}
\hline
&&Critical point \#9\\
\hline
$\phi_{01}$&=&-0.14838530863599789543465384360815968900455051258129946036956115829\\
   &&\phantom{+0.}6912659174620420013261089599470046207901483798616258605178379830\\
\hline
$\phi_{02}$&=&-0.29677061727199579086930768721631937800910102516259892073912231659\\
   &&\phantom{+0.}3825318349240840026522179198940092415802967597232517210356759659\\
\hline
$\phi_{03}$&=&-0.25732646426692985616707435046847694741517689249474313726724786427\\
   &&\phantom{+0.}2674663810312601338119308768781920952354455078216408698296166166\\
\hline
$\phi_{04}$&=&-0.21788231126186392146484101372063451682125275982688735379537341195\\
   &&\phantom{+0.}1524009271384362649716438338623749488905942559200300186235572673\\
\hline
$\phi_{05}$&=&-0.16737947827750216832085744976228405120097892770824596184306966143\\
   &&\phantom{+0.}4570173530882288314997735806904725069736607758864110903053338179\\
\hline
$\phi_{06}$&=&-0.11687664529314041517687388580393358558070509558960456989076591091\\
   &&\phantom{+0.}7616337790380213980279033275185700650567272958527921619871103685\\
\hline
$\phi_{07}$&=&-0.05843832264657020758843694290196679279035254779480228494538295545\\
   &&\phantom{+0.}8808168895190106990139516637592850325283636479263960809935551843\\
\hline
$\phi_{08}$&=&-0.08359029866858233034547941217571396802759842476707237180476077865\\
   &&\phantom{+0.}8414463573049930973956617579727777325402250717989610457138556824\\
\hline
$\phi_{11}$&=&+0.11635037784601938286230192233288778591954537037069254009954531733\\
   &&\phantom{+0.}0228235640944330444860413700610898200934371266006226890321421623\\
\hline
$\phi_{21}$&=&-0.09047325343654872016078886417723869587535769000540920521539879642\\
   &&\phantom{+0.}5492592428866196558540192536451998332250749011873992524792510049\\
\hline
$\phi_{30}$&=&+0.11529784295177731823315799539079618659722591993286848051710413015\\
   &&\phantom{+0.}5452031342072563374023174551461293301668567880962837431222057500\\
\hline
$\phi_{35}$&=&-0.10946742307805299304699247033136305807178610513235570668890729956\\
   &&\phantom{+0.}3150106785128064860276838743886677194085872972121844822667468398\\
\hline
$\phi_{36}$&=&+0.24667108042473646307996708755925237239714154731457784561932927971\\
   &&\phantom{+0.}9762167903809320049487455179596894551403218163281674880317791563\\
\hline
$\phi_{37}$&=&+0.49334216084947292615993417511850474479428309462915569123865855943\\
   &&\phantom{+0.}9524335807618640098974910359193789102806436326563349760635583126\\
\hline
$\phi_{38}$&=&+0.41111846737456077179994514593208728732856924552429640936554879953\\
   &&\phantom{+0.}2936946506348866749145758632661490919005363605469458133862985938\\
\hline
$\phi_{39}$&=&+0.32889477389964861743995611674566982986285539641943712749243903962\\
   &&\phantom{+0.}6349557205079093399316606906129192735204290884375566507090388750\\
\hline
$\phi_{43}$&=&-0.08222369347491215435998902918641745746571384910485928187310975990\\
   &&\phantom{+0.}6587389301269773349829151726532298183801072721093891626772597187\\
\hline
\end{longtable}
}



\end{document}